\documentclass[aps,prb,reprint,longbibliography,citeautoscript,superscriptaddress,showpacs]{revtex4-1}

\usepackage{graphicx,epsfig,float}
\usepackage{amssymb} 
\usepackage{amsmath}

\begin{document}
\title{Dislocations in stacking and commensurate-incommensurate phase transition in bilayer graphene and hexagonal boron nitride}

\author{Irina V. Lebedeva}
\email{liv\_ira@hotmail.com}
\affiliation{Nano-Bio Spectroscopy Group and ETSF, Universidad del Pa\'is Vasco, CFM CSIC-UPV/EHU, 20018 San Sebasti\'an, Spain
}
\author{Alexander V. Lebedev}
\email{allexandrleb@gmail.com}
\affiliation{Kintech Lab Ltd., 3rd Khoroshevskaya Street 12, Moscow 123298, Russia}
\author{Andrey M. Popov}
\email{popov-isan@mail.ru}
\affiliation{Institute for Spectroscopy of Russian Academy of Sciences, Troitsk, Moscow 142190, Russia}
\author{Andrey A. Knizhnik}
\email{kniznik@kintechlab.com}
\affiliation{Kintech Lab Ltd., 3rd Khoroshevskaya Street 12, Moscow 123298, Russia}
\affiliation{National Research Centre ``Kurchatov Institute", Moscow 123182, Russia}

\begin{abstract}
Dislocations corresponding to a change of stacking in two-dimensional hexagonal bilayers, graphene and boron nitride, and associated with boundaries between commensurate domains are investigated using the two-chain Frenkel-Kontorova model on top of \textit{ab initio} calculations. Structural transformations of bilayers in which the bottom layer is stretched and the upper one is left to relax freely are considered for gradually increased elongation of the bottom layer. Formation energies of dislocations, dislocation width and orientation of the boundary between commensurate domains are analyzed depending on the magnitude and direction of elongation. The second-order phase transition from the commensurate phase to the incommensurate one with multiple dislocations is predicted to take place at some critical elongation. The order parameter for this transition corresponds to the density of dislocations, which grows continuously upon increasing the elongation of the bottom layer above the critical value. In graphene and metastable boron nitride with the layers aligned in the same direction, where elementary dislocations are partial, this transition, however, is preceded by formation of the first dislocation at the elongation smaller than the critical one. The phase diagrams including this intermediate state are plotted in coordinates of the magnitude and direction of elongation of the bottom layer.
\end{abstract}

\maketitle

\section{Introduction}
\begin{figure*}
	\centering
	\includegraphics[width=\textwidth]{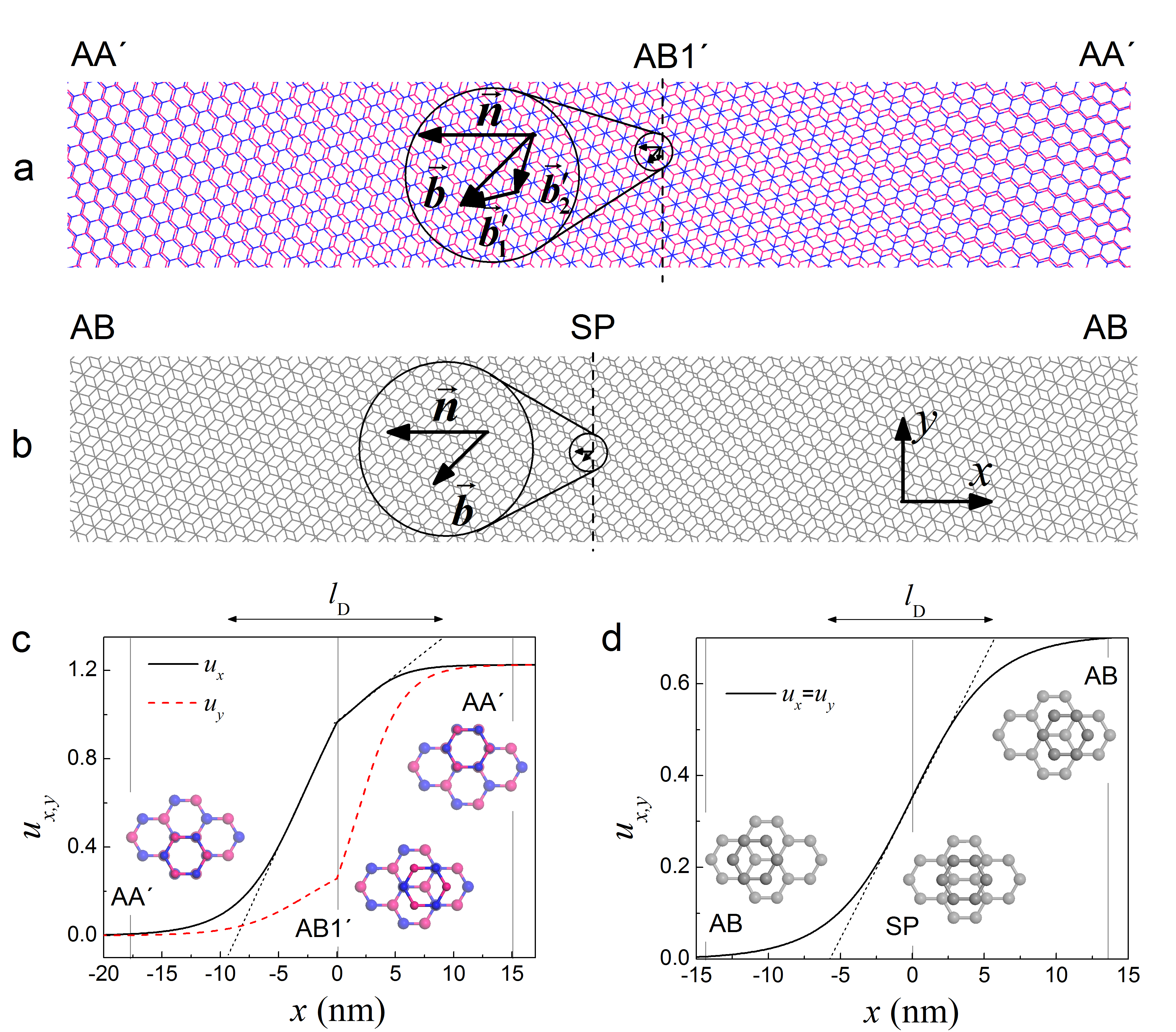}
	\caption{(Color online) (a,b) Atomistic structures corresponding to (a) a full dislocation in bilayer hexagonal boron nitride with the layers aligned in the opposite directions and (b) a partial dislocation in bilayer graphene with the angle $\beta = 45^{\circ}$ between the Burgers vector $\vec{b}$ and normal $\vec{n}$ to the boundary between commensurate domains. Boron, nitrogen and carbon atoms are coloured in blue/dark gray, magenta/ medium gray and light gray, respectively. The magnitude of the vector $\vec{b}$ is scaled up for clarity. (a) Vectors $\vec{b'}_1$ and $\vec{b'}_2$ corresponding to two straight parts of the dislocation path ($\vec{b} = \vec{b'}_1+\vec{b'}_2$) are shown.  (c,d) Displacements $\vec{u}$  (in units of bond length $l$) of atoms as functions of their position $x$ (in nm) in the direction of the normal $\vec{n}$ to the boundary between commensurate domains (a) for the full dislocation in hexagonal boron nitride with the layers aligned in the opposite directions and (b) for the partial dislocation in graphene. The displacements perpendicular to the boundary between commensurate domains ($u_x$, black solid lines) and along the boundary ($u_y$, red dashed lines) are shown (the two curves are the same for the considered partial dislocation in graphene). The atomistic structures of symmetric stackings across the boundaries between commensurate domains are included with their position shown by vertical gray lines. The characteristic widths $l_D$ of the dislocations are indicated.}
	\label{fig:disl_struct}
\end{figure*}

Dislocations are common defects in any crystals \cite{Chaikin1995}. They are not only responsible for plastic deformation of the materials but also affect their electronic, optical \cite{Pennycook2008,Shreter1993} and transport properties \cite{Holt2007,Chernatynskiy2012}. Two-dimensional crystals based on novel layered materials, such as graphene and hexagonal boron nitride, are not exceptions \cite{Alden2013,Butz2014,Lin2013,Yankowitz2014}. In this case, dislocations can be divided into two very distinct classes: dislocations within the layers disturbing perfect hexagonal ordering of atoms (see review \cite{Skowron2015}) and dislocations between the layers manifested through the changes in the layer stacking \cite{Popov2011,Lebedev2015}. The latter ones (Fig.~\ref{fig:disl_struct}), though often observed in the experimental images, where the bilayer or few-layer system appears divided into a number of commensurate domains separated by incommensurate boundaries \cite{Alden2013,Butz2014,Lin2013,Yankowitz2014}, are much less studied compared to the dislocations within the layers and other in-plane structural defects \cite{Skowron2015}. Nevertheless, numerous experimental and theoretical studies show that such dislocations tune electronic \cite{Hattendorf2013, San-Jose2014, Lalmi2014, Benameur2015, Koshino2013} and optical \cite{Gong2013} properties of graphene and thus can be used for development of nanoelectronic devices. In the present paper, we use the formalism of the Frenkel-Kontorova model \cite{Chaikin1995,Bichoutskaia2006,Popov2011} on top of \textit{ab initio} calculations to analyze structure and energetics of dislocations in stacking of two-dimensional bilayers.  

The presence of two degenerate but inequivalent minima AB and AC on the potential surface of interlayer interaction energy of bilayer graphene provides that elementary dislocations in stacking in this material are partial (Fig.~\ref{fig:disl_struct}b)  \cite{Popov2011,Alden2013,Butz2014,Lin2013,Yankowitz2014}. The Burgers vector $\vec{b}$ of such dislocations is equal in magnitude to the bond length $l$ and is smaller than the lattice constant $a_0 = l \sqrt{3}$. We suggest that partial dislocations can be also observed in hexagonal boron nitride with co-aligned layers (AB stacking in the commensurate state) since the symmetry of its potential energy surface is the same as the one for graphene \cite{Lebedev2015}. Though such a stacking mode of hexagonal boron nitride layers does not correspond to the global energy minimum, it is metastable and has been observed experimentally \cite{Warner2010}. Therefore, we consider formation of partial dislocations both in bilayer graphene and hexagonal boron nitride with the layers aligned in the same direction. In boron nitride with the layers aligned in opposite directions, on the other hand, two non-equivalent minima on the potential energy surface are rather different in energy. Thus, in this material, we study full dislocations with the Burgers vector $b = l \sqrt{3}$ (Fig.~\ref{fig:disl_struct}a) \cite{Lebedev2015}.

The formation energy of dislocations, which includes contributions to the elastic energies of the layers and the energy of their interaction and is proportional to the length of the boundary between commensurate domains, can be significant in the absence of external strains \cite{Popov2011,Alden2013,Lebedev2015}. Application of external strains to one of the layers, however, changes the balance of the elastic and interlayer interaction energies \cite{Popov2011}. At small strains, the interlayer interaction tends to keep the layers commensurate. Upon increasing the strain to some critical value it becomes favourable to release the excessive elastic energy by formation of a dislocation. Increasing strains further leads to generation of more and more dislocations. In the limit of infinite length of the system the density of dislocations changes continuously upon increasing the strain above the critical value and the second-order phase transition from the commensurate to incommensurate phase characterized by the density of dislocations as the order parameter takes place \cite{Pokrovsky1978,Popov2011}. Commensurate and incommensurate phases have been widely studied for adsorbates and commensurate-incommensurate phase transitions have been observed upon change in temperature, pressure or coverage \cite{Chaikin1995}. The discovery of one-dimensional crystals such as carbon nanotubes and two-dimensional crystals such as graphene and boron nitride make it possible to observe the commensurate-incommensurate phase transition by stretching of one of the initially commensurate layers \cite{Popov2011} or commensurate walls \cite{Bichoutskaia2006,Popov2009}. The crossover from the state with commensurate domains separated by incommensurate boundaries to the fully incommensurate state was already observed for the layers with a small lattice constant mismatch by changing the relative orientation of the layers \cite{Woods2014}.

\begin{figure}
	\centering
	\includegraphics[width=\columnwidth]{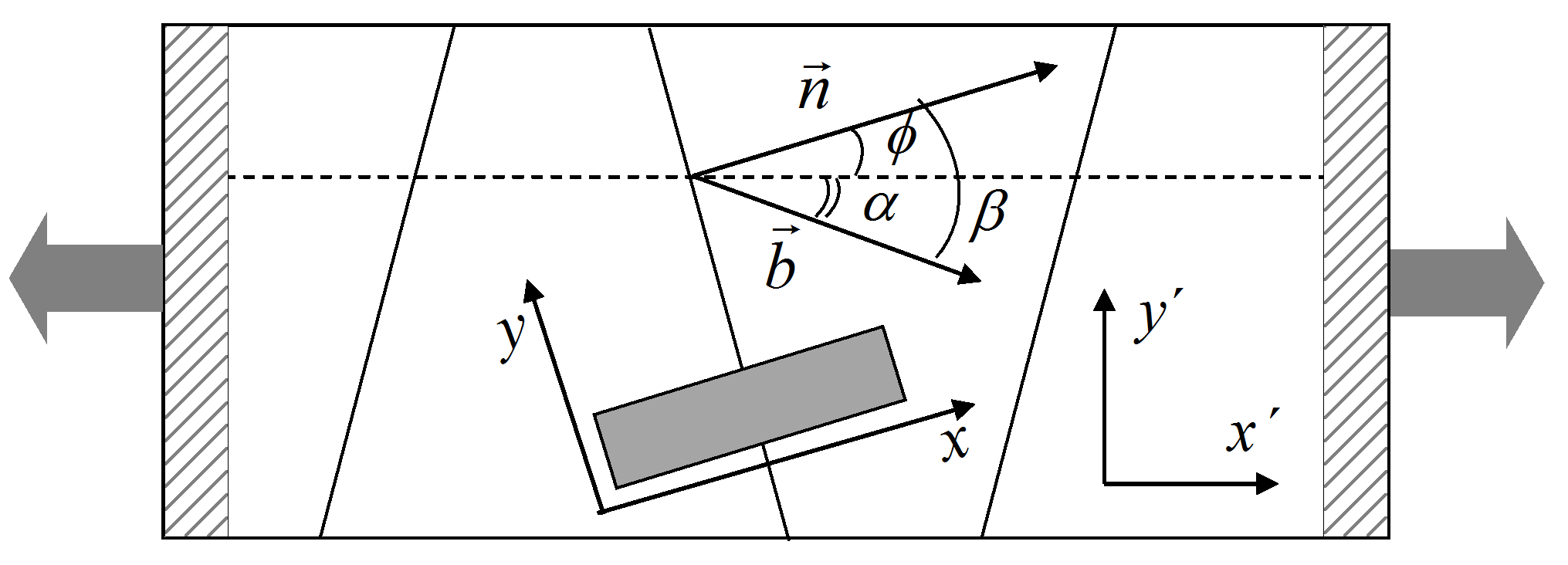}
	\caption{Scheme of generation of partial dislocations (inclined lines) in bilayer graphene or hexagonal boron nitride with the layers aligned in the same direction upon stretching the bottom layer (shaded area). The angles characterizing relative orientation of the normal $\vec{n}$ to the boundary between commensurate domains, Burgers vector $\vec{b}$ of the dislocation and direction of elongation of the bottom layer (dashed line) are indicated. $\alpha$ is the angle between the direction of elongation and Burgers vector $\vec{b}$. $\phi$ is the angle between the normal $\vec{n}$ to the boundary between commensurate domains and direction of elongation. $\beta = \alpha + \phi$ is the angle between the normal $\vec{n}$ to the boundary between commensurate domains and Burgers vector $\vec{b}$. The coordinate systems $x-y$ and $x'-y'$ associated with a single dislocation and elongation applied, respectively, are shown.} 
	\label{fig:statement}
\end{figure}

In the present paper we consider graphene and boron nitride bilayers with the gradually stretched bottom layer and free upper layer (Fig.~\ref{fig:statement}). The potential energy surfaces of bilayer graphene and hexagonal boron nitride obtained on the basis of density functional theory (DFT) calculations are used to estimate characteristics of dislocations in these materials. Phase diagrams for the commensurate-incommensurate phase transition in coordinates of the magnitude and direction of elongation are obtained. Differences in structural transformations taking place upon stretching the bottom layer are revealed in boron nitride with the layers aligned in the opposite directions, where the elementary dislocations are full, and graphene and boron nitride with the layers aligned in the same direction, where the elementary dislocations are partial.

The estimated physical quantities can be accessed experimentally, for example, by transmission electron microscopy. Comparison of the theoretical estimates and experimental data could provide a valuable insight into interaction of atomically thin layers. Experimental observations of pre-existing dislocations in few-layer graphene already allowed to get an experimental estimate of the barrier to relative sliding of graphene layers \cite{Alden2013}.
Though in the present paper we focus on structural transformations that take place in two-dimensional layered materials upon stretching (Fig.~\ref{fig:statement}), similar phenomena can be observed upon shear or bending deformations \cite{Korhonen2015}.   

The paper is organized in the following way. In section 2 we discuss basic characteristics of interlayer interaction and elastic properties of graphene and boron nitride layers. In section 3 the energetics and structure of dislocations in bilayers are analyzed depending on the elongation of the bottom layer and the commensurate-incommensurate phase transition is considered. Finally conclusions are summarized.

\section{Interlayer interaction and elastic properties}
The shape and energy of dislocations in stacking of two-dimensional layers are determined by the balance of elastic energies of the deformed layers and the energy of their interaction. In the present paper the elastic properties and potential energy surfaces, i.e. dependences of the interlayer interaction energy on the relative position of the layers, are obtained by DFT calculations taking into account van der Waals interactions through the non-local vdW-DF2 functional \cite{Lee2010}. The calculations are performed using VASP code \cite{Kresse1996}. The maximum kinetic energy of plane waves is 600 eV. The interaction of valence electrons is described using the projector augmented-wave method (PAW) \cite{Kresse1999}. The convergence threshold of the self-consistent field is $10^{-6}$ eV. The rectangular unit cell with 4 atoms in each layer and height of 20~\AA~is considered under periodic boundary conditions. Integration over the Brillouin zone is performed using the Monkhorst-Pack method \cite{Monkhorst1976} with at least $24\times 20 \times 1$ k-points (in the armchair and zigzag directions, respectively). These parameters provide convergence of the barriers to relative sliding of the layers with respect to the number of k-points and maximum kinetic energy of plane waves within 2\% \cite{Lebedeva2011}.  

The optimized bond lengths for graphene and hexagonal boron nitride are $l = 1.430$~\AA~and 1.455~\AA, respectively, close to the experimental data for graphite \cite{Bernal1924, Baskin1955, Wyckoff1963,  Lynch1966, Ludsteck1972, Trucano1975, Zhao1989, Bosak2007} and bulk boron nitride \cite{Pease1950, Pease1952, Lynch1966, Solozhenko1995, Solozhenko1997, Solozhenko2001, Paszkowicz2002, Bosak2006, Fuchizaki2008}. The potential surface of interlayer interaction energy is considered at a fixed interlayer distance known from the experiments of $d=3.33$~\AA~ for bulk boron nitride \cite{Pease1950, Pease1952, Lynch1966, Solozhenko1995, Solozhenko1997, Solozhenko2001, Paszkowicz2002, Bosak2006, Fuchizaki2008} and $d=3.34$~\AA~for graphene \cite{Bernal1924, Baskin1955, Wyckoff1963, Lynch1966, Ludsteck1972, Trucano1975, Zhao1989, Bosak2007}. Setting the interlayer distance at the experimental value combined with the use of vdW-DF2 functional \cite{Lee2010} allows to get the potential energy surfaces \cite{Lebedev2015} in close agreement with the local second-order M{\o}ller-Plesset perturbation theory (LMP2) \cite{Constantinescu2013}, which is a high-level \textit{ab initio} method that adequately describes van der Waals interactions.

\begin{figure*}
	\centering
	\includegraphics[width=\textwidth]{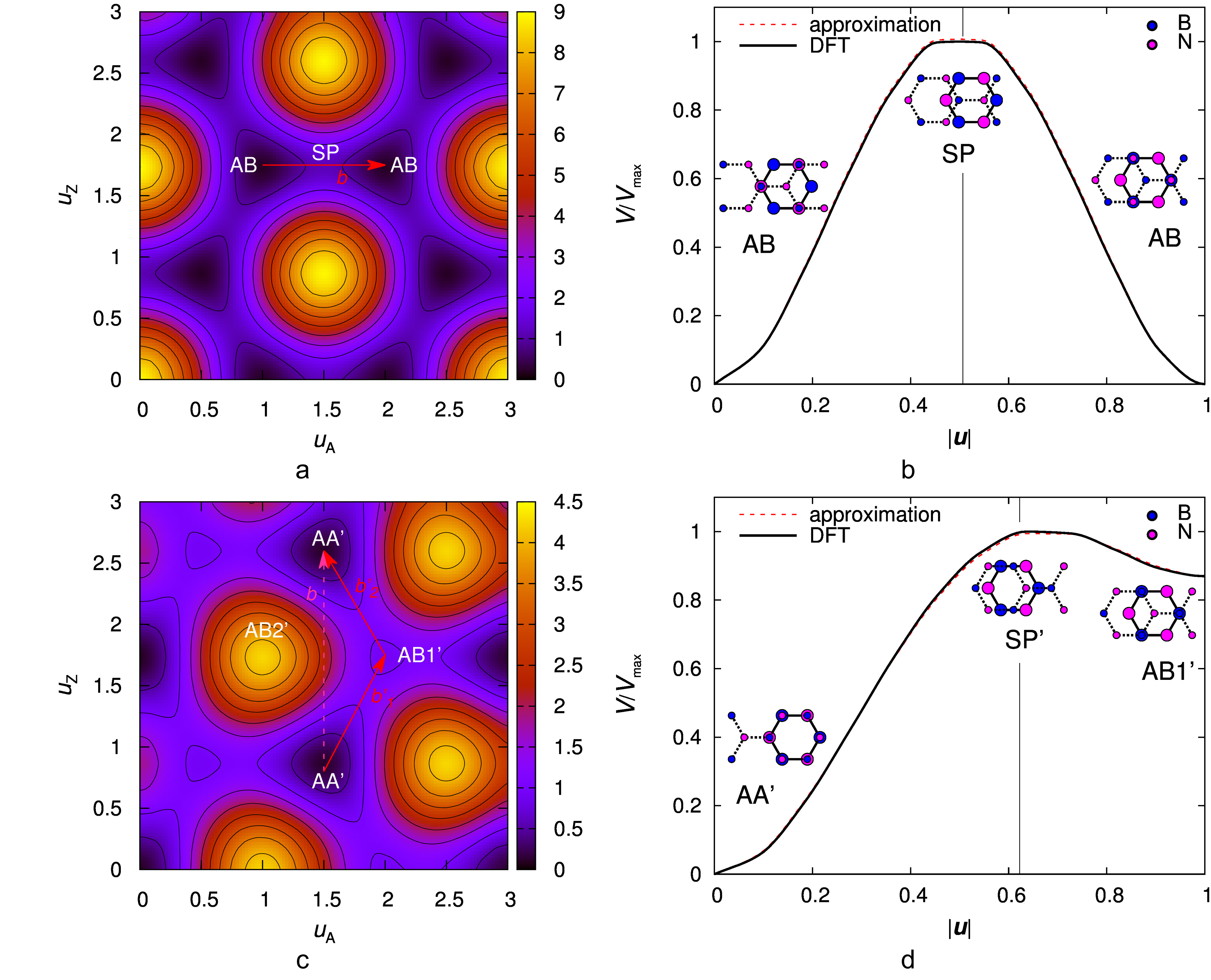}
	\caption{(Color online) Calculated interlayer interaction energy of bilayer hexagonal boron nitride $V$ (in units of the barrier $V_\mathrm{max}$ to relative sliding of the layers) as a function of relative displacement of the layers in the armchair ($u_A$, in units of bond length $l$) and zigzag ($u_Z$, in units of bond length $l$) directions at the interlayer distance of $d=$3.33~\AA: (a,b) layers aligned in the same direction and (c,d) layers aligned in the opposite directions. The energy is given relative to the AB (a,b) or AA' (c,d) stacking, respectively. The Burgers vectors corresponding to the (c) full ($b = l\sqrt{3}$) and (a) partial ($b = l$) dislocations are indicated. (c) The vectors $\vec{b'}_1$ and $\vec{b'}_2$ corresponding to two straight parts of the dislocation path ($\vec{b} = \vec{b'}_1+\vec{b'}_2$) are shown. (b,d) Black solid lines correspond to the calculated dependences of interlayer interaction energy $V/V_\mathrm{max}$ on displacement $u$ along the minimum energy paths indicated in figures (a,c). Curves approximated according to Eqs.~(\ref{eq_a1}) (b) and (\ref{eq_a2}) (d) are shown by red dashed lines. Structures of the symmetric stackings are indicated. Boron and nitrogen atoms are coloured in blue/dark gray and magenta/medium gray, respectively.}
	\label{fig:pes}
\end{figure*}

The calculated potential energy surfaces for bilayer boron nitride with the layers aligned in the same or opposite directions are presented in Fig.~\ref{fig:pes}. It should be noted that for graphene and bilayer boron nitride with the layers aligned in the same direction (Fig.~\ref{fig:pes}a), the potential energy surfaces are very similar when they are expressed in the units of the bond length $l$ and barrier to relative sliding of the layers $V_\mathrm{max}$ \cite{Lebedev2015}. The minima on the potential energy surface in this case correspond to the AB stacking in which half of atoms of the upper layer are located on top of atoms of the bottom layer and the other half on top of centers of hexagons (Fig.~\ref{fig:pes}b) \cite{Lebedev2015,Popov2012,Kolmogorov2005,Reguzzoni2012,Aoki2007,Ershova2010,Lebedeva2011,Lebedeva2010,Lebedeva2011a}.
The barrier to relative sliding of the layers is determined by the energy of the saddle-point stacking (SP) stacking relative to the AB stacking. The barrier calculated for graphene layers is $V_\mathrm{max} = 1.61$~meV/atom (note that all energies for bilayers in the present paper are given in meV per atom in the upper (adsorbed) layer). This value is in agreement with the experimental estimates from the shear mode frequency and width of dislocations of 1.7 meV/atom~\cite{Popov2012} and 2.4 meV/atom~\cite{Alden2013}, respectively. It is also within the range of DFT values of 0.5 -- 2.1 meV/atom reported previously for bilayer graphene \cite{Dion2004,Aoki2007,Lebedeva2011,Reguzzoni2012}, graphite \cite{Kolmogorov2005} and polycyclic aromatic hydrocarbons adsorbed on graphene \cite{Ershova2010}. The calculated barrier to relative sliding of co-aligned boron nitride layers of  $V_\mathrm{max}=1.92$~meV/atom is close to the LMP2 result of $\sim$2.5~meV/atom \cite{Constantinescu2013}.

For boron nitride layers aligned in the opposite directions, there are two non-degenerate energy minima \cite{Lebedev2015,Marom2010,Constantinescu2013,Gao2015} (Fig.~\ref{fig:pes}c). The most stable ones correspond to the AA' stacking \cite{Lebedev2015,Constantinescu2013,Gao2015} with boron (nitrogen) atoms of the upper layer on top of nitrogen (boron) atoms of the bottom layer (Fig.~\ref{fig:pes}d). The shallow minima correspond the AB1' stacking with boron atoms of the upper layer on top of boron atoms of the bottom layer and the rest of atoms of the upper layer on top of hexagon centers. According to our calculations, this stacking is $E_{AB1'} = 3.10$ meV/atom less stable than the AA' stacking, in agreement with the LMP2 result of 4.4 meV/atom \cite{Constantinescu2013}. The calculated barrier to relative sliding of the layers is $V_\mathrm{max} = 3.57$~meV/atom.

To calculate the formation energy of dislocations it is needed to know the interlayer interaction energy along the dislocation path, i.e. along the curve on the potential energy surface described by the dependence of relative displacement $\vec{u}$ of the layers on the coordinate $x$ in the direction perpendicular to the boundary between commensurate domains that minimizes the formation energy of dislocations. As discussed in Appendix, the dislocation path of partial dislocations in graphene or hexagonal boron nitride with the layers aligned in the same direction exactly corresponds to the minimum energy path between two adjacent minima and is represented by a straight line AB -- AB in one of the armchair directions (Figs.~\ref{fig:pes}a and b). In the case of full dislocations in boron nitride layers aligned in the opposite directions, we suggest that the dislocation path can also be approximated by the minimum energy path consisting of two straight lines AA' -- AB1' -- AA' (Figs.~\ref{fig:pes}c and d). 

It has been shown in the previous papers that the potential energy surfaces of graphene \cite{Lebedeva2010,Lebedeva2011a,Popov2012,Reguzzoni2012} and hexagonal boron nitride \cite{Lebedev2015} can be described with high accuracy by expressions containing only the first Fourier harmonics. In the case of graphene layers  \cite{Lebedeva2010,Lebedeva2011a,Popov2012,Reguzzoni2012} or boron nitride layers aligned in the same direction \cite{Lebedev2015}, this means that the dependence of the interlayer interaction energy $V$ on the relative displacement $u$ of the layers along the dislocation path AB -- AB of partial dislocations (Fig.~\ref{fig:pes}b) can be represented as 
\begin{equation} \label{eq_a1}
\begin{split}
	V(u) = V_\mathrm{max}\left(2\cos{\left(k_0 u + \frac{2\pi}{3}\right)} +1\right)^2,
\end{split}
\end{equation}
where $k_0 = 2\pi/3$ (Fig.~\ref{fig:pes}a) and $u$ is in units of the bond length $l$, $0 \le u \le 1$. As shown in the next section, this minimum energy path provides contribution $\int_0^1 \sqrt{V(u)} \mathrm{d} u = 0.654 \sqrt{V_\mathrm{max}}$ to the formation energy of dislocations.

Taking into account the relative energy of the AB1' stacking $E_{AB1'}/V_\mathrm{max} = 0.870$, the variation of the interaction energy $V$ along the straight piece of the dislocation path AA' -- AB1' for boron nitride layers aligned in the opposite directions \cite{Lebedev2015} can be approximated as (Fig.~\ref{fig:pes}d)
\begin{equation} \label{eq_a2}
\begin{split}
  V(u)&= V_\mathrm{max} \Big\{ 2.646 - 0.294 \times (2\cos{(k_0 u)} + 1)^2\\
     &- 2.734 \times \sin{(k_0 u)} \sin^2{\left(k_0 u/2\right)} \Big\},
\end{split}
\end{equation}
giving $\int_0^1 \sqrt{V(u)} \mathrm{d} u =0.764 \sqrt{V_\mathrm{max}}$. 
 
To estimate the Young modulus and Poisson ratio strains up to 0.8\% in the armchair direction and up to 0.5\% in the perpendicular zigzag direction have been applied to single layers of graphene and boron nitride. Positions of atoms within the cell are optimized using the quasi-Newton method \cite{Pulay1980} till the residual force of $6 \cdot 10^{-3}$~eV/\AA. For each value of strain in the armchair direction, the energy as a function of strain in the zigzag direction is approximated by a parabola. The elastic constant is found by the dependence of the energy corresponding to the parabola minimum on strain in the armchair direction. The Poisson ratio is determined by the dependence of the position of this minimum on strain. For hexagonal boron nitride, our calculations give the elastic constant $k = 272.8 \pm 0.5$~J/m$^2$ or Young modulus $Y = k/d = 819 \pm 2$~GPa and the Poisson ratio $\nu = 0.201 \pm 0.001$, in excellent agreement with the experimental data for bulk of $Y = 811 \pm 12$~GPa and $\nu = 0.21 \pm 0.03$ (Ref.~\onlinecite{Bosak2006}). For graphene, the calculated values are $k = 331 \pm 1$~J/m$^2$, $Y = k/d = 991 \pm 4$~GPa and $\nu = 0.174 \pm 0.002$, in agreement with the experimental data for graphite of $Y = 1109 \pm 16$~GPa and $\nu = 0.13 \pm 0.03$~(Ref.~\onlinecite{Bosak2007}), $Y = 1060 \pm 20$~GPa and $\nu = 0.17 \pm 0.02$~(Ref.~\onlinecite{Blakslee1970}) and for single-layer graphene of $k = 340 \pm 50$~J/m$^2$~(Ref.~\onlinecite{Lee2008}). 

\section{Dislocations and phase transition}
\subsection{Model description}
Let us now consider the energy and structure of dislocations in stacking of bilayer graphene and boron nitride  (Fig.~\ref{fig:disl_struct}). To include the effect of external strains and analyze the commensurate-incommensurate phase transition we study the case when the bottom layer is stretched, while the upper layer is left to relax freely (Fig.~\ref{fig:statement}). The direction of elongation of the bottom layer is arbitrary. However, we assume that distances between boundaries separating commensurate domains exceed considerably their widths and there are no boundary crossings. This is a reasonable assumption for samples with the high aspect ratio and low density of dislocations. It is also supposed that the sample width is much greater than the width of boundaries between commensurate domains so that the edge effects can be neglected.

Our derivations are based on the formalism of the two-chain Frenkel-Kontorova model \cite{Bichoutskaia2006}, which was successfully applied previously to study dislocations in double-walled carbon nanotubes \cite{Bichoutskaia2006, Popov2009} and the tensile partial dislocation in graphene \cite{Popov2011}. In this model two chains of particles connected by harmonic springs are considered. The chains are coupled through the forces corresponding to the van der Waals interaction of the layers. To apply this model to two-dimensional layers it is assumed that the particle-spring pairs correspond to ribbons of the layers parallel to the boundary between commensurate domains \cite{Popov2011}. Since the standard model \cite{Bichoutskaia2006, Popov2009,Popov2011} is purely one-dimensional it is limited to tensile dislocations under tensile elongation \cite{Popov2011}. We extend the model to systems where atoms of the layers are displaced on the two-dimensional surface, the Burgers vector forms an arbitrary angle to the boundary between commensurate domains and arbitrary strain is applied. 
In this case springs should be substituted by two-dimensional objects which elastic properties are described by the elasticity tensor. Then it is easier to skip the step of discrete particles and to start directly from the continuum formulation of the model. It should be noted, however, that we still restrict ourselves to consideration of isolated dislocations where relative displacements of atoms of the layers are functions only of the coordinate $x$ across the boundary between commensurate domains and do not depend on the coordinate $y$ along the boundary (Fig.~\ref{fig:statement}, we consider the coordinate system of the gray rectangular). 

To obtain the energy per unit length of the boundary between commensurate domains we represent it as $W = \int w \mathrm{d}x$, where $w$ is the energy density given by the sum $w=w_\mathrm{el}^{(1)} +w_\mathrm{el}^{(2)}+w_\mathrm{int}$ of densities of elastic energies of the layers $w_\mathrm{el}^{(1,2)}$ and their interaction energy $w_\mathrm{int}$. Let us first consider the elastic energies. The tensile strain $\epsilon_{i,x}$ in the direction $x$ across the boundary between commensurate domains and shear strain $\gamma_{i,xy}$ in layer $i$ include external strains $\epsilon_x$ and $\gamma_{xy}$ and strain $\vec{v}^{\prime}_i =\mathrm{d} \vec{v}_i /\mathrm{d}x$ associated with formation of the dislocation, where $\vec{v}_i(x)$ are displacements of atoms relative to their regular positions in the commensurate bilayer when only uniform external strains are present (analogous to displacements of particles of the one-dimensional Frenkel-Kontorova model relative to equidistant positions in the commensurate chains). Therefore, $\epsilon_{i,x}=\epsilon_{x}+ v_{i,x}^{\prime}$ and $\gamma_{i,xy}=\gamma_{xy}+ v_{i,y}^{\prime}$, respectively.  Since the layers are commensurate in the direction $y$ along the boundary, tensile strains in this direction are the same for both of the layers and are equal to the external strain $\epsilon_{i,y}=\epsilon_{y}$. The tensile and shear stresses in the layers, however, are different for all the directions and are related to the strains through the elastic constant under uniaxial stress  $k = Yd$ and Poission ratio $\nu$ as $\sigma_{i,x}=E(\epsilon_{i,x}+\nu\epsilon_{i,y})$, $\sigma_{i,y}=E(\epsilon_{i,y}+\nu\epsilon_{i,x})$ and $\tau_{i,xy}=G\gamma_{i,xy}$, where $E = k/(1-\nu^2)$ and $G = k/2(1+\nu)$ are the tensile and shear elastic constants per unit area, respectively. Using these strains and stresses, the density of elastic energy of layer $i$ can be found as
\begin{equation} \label{eq_1}
\begin{split}
w_\mathrm{el}^{(i)} = &
\frac{1}{2} \left(\epsilon_{i,x} \sigma_{i,x} +\epsilon_{i,y} \sigma_{i,y}   + \gamma_{i,xy} \tau_{i,xy} \right) .
\end{split}
\end{equation}
The density of the interaction energy is $w_\mathrm{int} = V(\vec{u})$, where $\vec{u} = (\vec{v}_1 - \vec{v}_2)/l$ is the relative displacement of the layers in units of the bond length $l$. 

It is also necessary to mention the boundary conditions for $\vec{v}_i$. The external strains are introduced through stretching or shear deformation the bottom layer so that the geometry of the bottom layer is fixed. This provides the boundary condition for the bottom layer $\Delta v_{2,x}  =  \Delta v_{2,y} = 0$, where $\Delta \vec{v}_{2} = \vec{v}_2 (+\infty) - \vec{v}_2 (-\infty)$. The upper layer is allowed to relax freely. 

In terms of variables $\vec{u}$ and $\vec{a} = (\vec{v}_1 + \vec{v}_2)/2l$, the energy of the bilayer with a dislocation relative to the commensurate system can be finally presented as $\Delta W = \Delta W_0 +\Delta W_\epsilon$, where $\Delta W_0$ is the energy in the absence of external strains and $\Delta W_\epsilon$ is the energy change provided by the external load. The first one is given by  
\begin{equation} \label{eq_2}
\begin{split}
\Delta W_0 = & \int\limits_{-\infty}^{+\infty} \bigg\{\frac{1}{4} K(\vec{u}^{\prime}) l^2 |\vec{u}^{\prime}|^2  + V(\vec{u})\bigg\}\mathrm{d}x \\
&+ \int\limits_{-\infty}^{+\infty} K(\vec{a}^{\prime}) l^2 |\vec{a}^{\prime}|^2 \mathrm{d}x.
\end{split}
\end{equation}
Here $K(\vec{u}^{\prime}) = E \cos^2{\theta (\vec{u}')} + G \sin^2  {\theta(\vec{u}')}$ describes the dependence of the elastic constant on fractions of tensile and shear character in the dislocation and $\theta$ is the angle between the dislocation path and normal to the boundary between commensurate domains $\sin{\theta} =u'_y/|\vec{u}'|$. 

In the presence of the external load, formation of dislocations can reduce the elastic energy of the free layer. The corresponding contribution to the formation energy of dislocations does not depend on the dislocation path
\begin{equation} \label{eq_2a}
\begin{split}
\Delta W_\epsilon =  El (\epsilon_x + \nu \epsilon_y) \Delta u_x  + Gl \gamma_{xy}  \Delta u_y.
\end{split}
\end{equation}
Here $\Delta \vec{u} = \vec{u} (+\infty) - \vec{u} (-\infty) = - \vec{b}/l$, $\vec{b}$ is the Burgers vector of the dislocation and we take into account the boundary conditions for the bottom layer.

\subsection{Formation energy and width of dislocations}
As discussed in Appendix, the dislocation path, i.e. the curve described by the dependence $\vec{u}(x)$ that minimizes the formation energy (Eqs.~(\ref{eq_2}) and (\ref{eq_2a})), lies along the minimum energy path AB -- AB for partial dislocations (Figs.~\ref{fig:pes}a and b) and approximately along the minimum energy path AA' -- AB1' -- AA'  (Figs.~\ref{fig:pes}c and d) for full dislocations. The dislplacement along these curves is driven by the analogue of the energy conservation law (obtained by integration of the Euler-Lagrange equations (\ref{eq_3}))
\begin{equation} \label{eq_3a}
\begin{split}
& \frac{1}{4} K(\vec{u}^{\prime}) l^2 |\vec{u}^{\prime}|^2 = V(\vec{u}) + c_1,\\
& K(\vec{a}^{\prime}) l^2 |\vec{a}^{\prime}|^2  = c_2.
\end{split}
\end{equation}
The integration constants $c_1$ and $c_2$ are associated with the energy cost for interaction of adjacent dislocations and determine the number of dislocations generated at high strains \cite{Chaikin1995}. In the limit of low density of dislocations and large systems \cite{Chaikin1995,Popov2011}, however, these integration constants tend to zero and can be neglected.  Using the equality in the density of the elastic and interaction energies of the layers that holds in this case in Eq.~(\ref{eq_2}), the formation energy of dislocations in the absence of external strains can be represented as the geometrical mean of the elastic and interaction energies 
\begin{equation} \label{eq_4}
\begin{split}
\Delta W_0 = \int\limits_{-\infty}^{+\infty} \sqrt{ K(\vec{u}^{\prime}) l^2 |\vec{u}^{\prime}|^2  V(\vec{u})}\mathrm{d}x.
\end{split}
\end{equation} 

\begin{figure}
	\centering
	\includegraphics[width=\columnwidth]{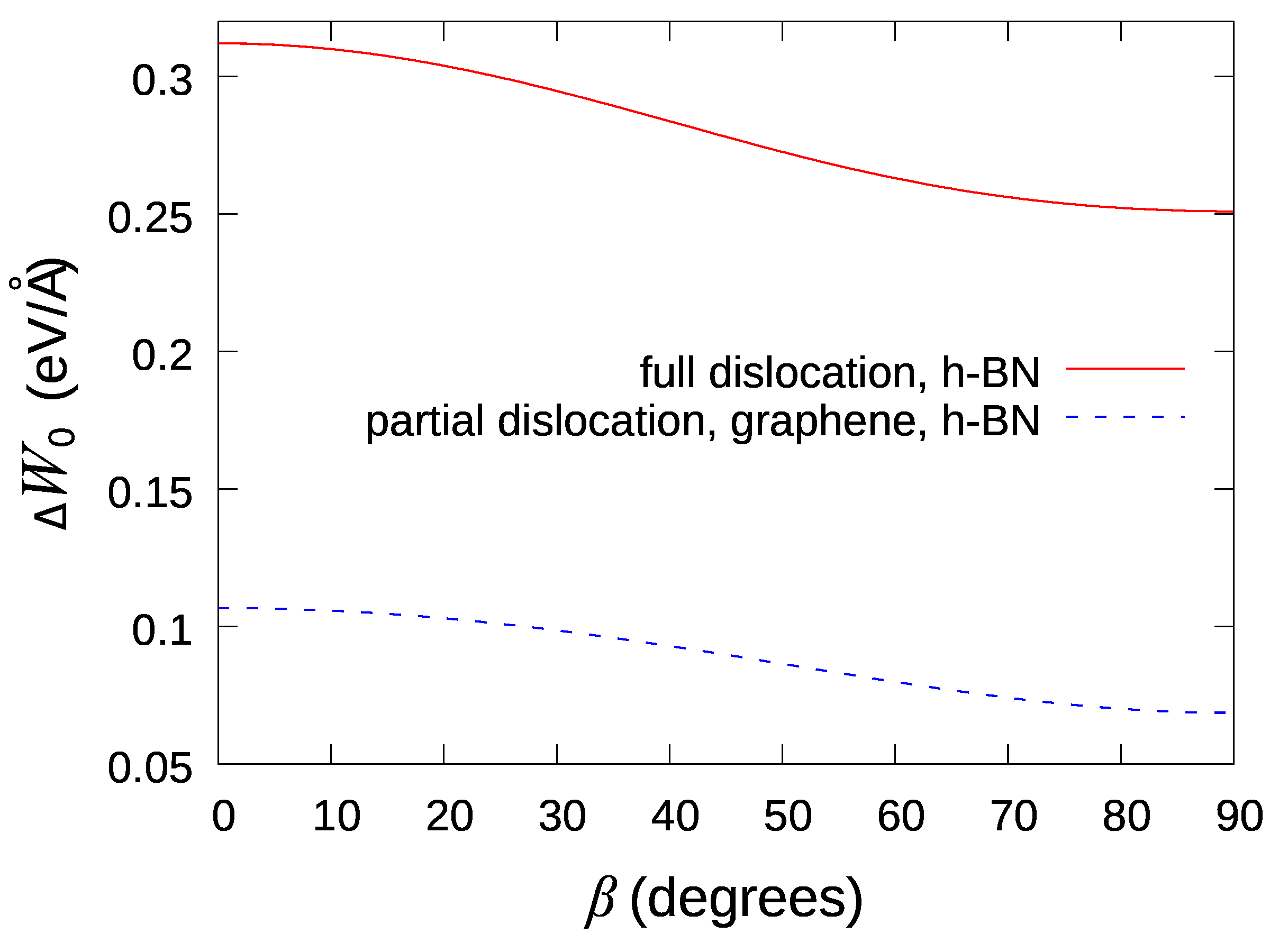}
	\caption{(Color online) Calculated formation energy of dislocations $\Delta W_0$ in the absence of external strains per unit length of the boundary between commensurate domains (in eV/\AA) as a function of angle $\beta$ (in degrees) between the Burgers vector $\vec{b}$ and normal $\vec{n}$ to the boundary between commensurate domains (Fig.~\ref{fig:statement}) for a full dislocation (red solid line) in bilayer hexagonal boron nitride with the layers aligned in the opposite directions and partial dislocations in bilayer graphene and boron nitride with the layers aligned in the same direction (blue dashed line). The curves for partial dislocations in bilayer graphene and boron nitride with the layers aligned in the same direction are too close to be distinguished in the figure (see text).}
	\label{fig:energy}
\end{figure}

Let us start from consideration of partial dislocations in graphene (Fig.~\ref{fig:disl_struct}b) and boron nitride layers aligned in the same direction. In these dislocations, the layers are displaced along straight lines in the direction opposite to the Brugers vector (Fig.~\ref{fig:pes}a), i.e. at the angle $\theta = 180^{\circ} + \beta$ to the normal to the boundary between commensurate domains, where $\beta$ is the angle between the Burgers vector and the normal to the boundary (Fig.~\ref{fig:statement}), while $u$ changes from 0 to 1. Then the formation energy of partial dislocations in the limit of zero external strain given by Eq.~(\ref{eq_4}) can be presented as $\Delta W_0 (\beta) =W_0 \sqrt{ K(\beta)/k}$ (Fig.~\ref{fig:energy}), where we denote $W_0 =\sqrt{kl^2}\int_0^1 \sqrt{V(u)} \mathrm{d} u$. It should be noted that the parameter $W_0$ takes very close values for bilayer graphene and boron nitride with the layers aligned in the same direction: 0.10505 eV/\AA~and 0.10532 eV/\AA, respectively. The formation energy of partial dislocations per unit length of the boundary between commensurate domains changes in the absence of external strains from $\Delta W_0 (90^{\circ}) = W_0/\sqrt{2(1+\nu)}$ for shear dislocations to $\Delta W_0 (0^{\circ}) =  W_0/\sqrt{1-\nu^2}$ for tensile dislocations, i.e. for the boundaries between commensurate domains in the armchair and zigzag directions, respectively.  

\begin{figure}
	\centering
	\includegraphics[width=\columnwidth]{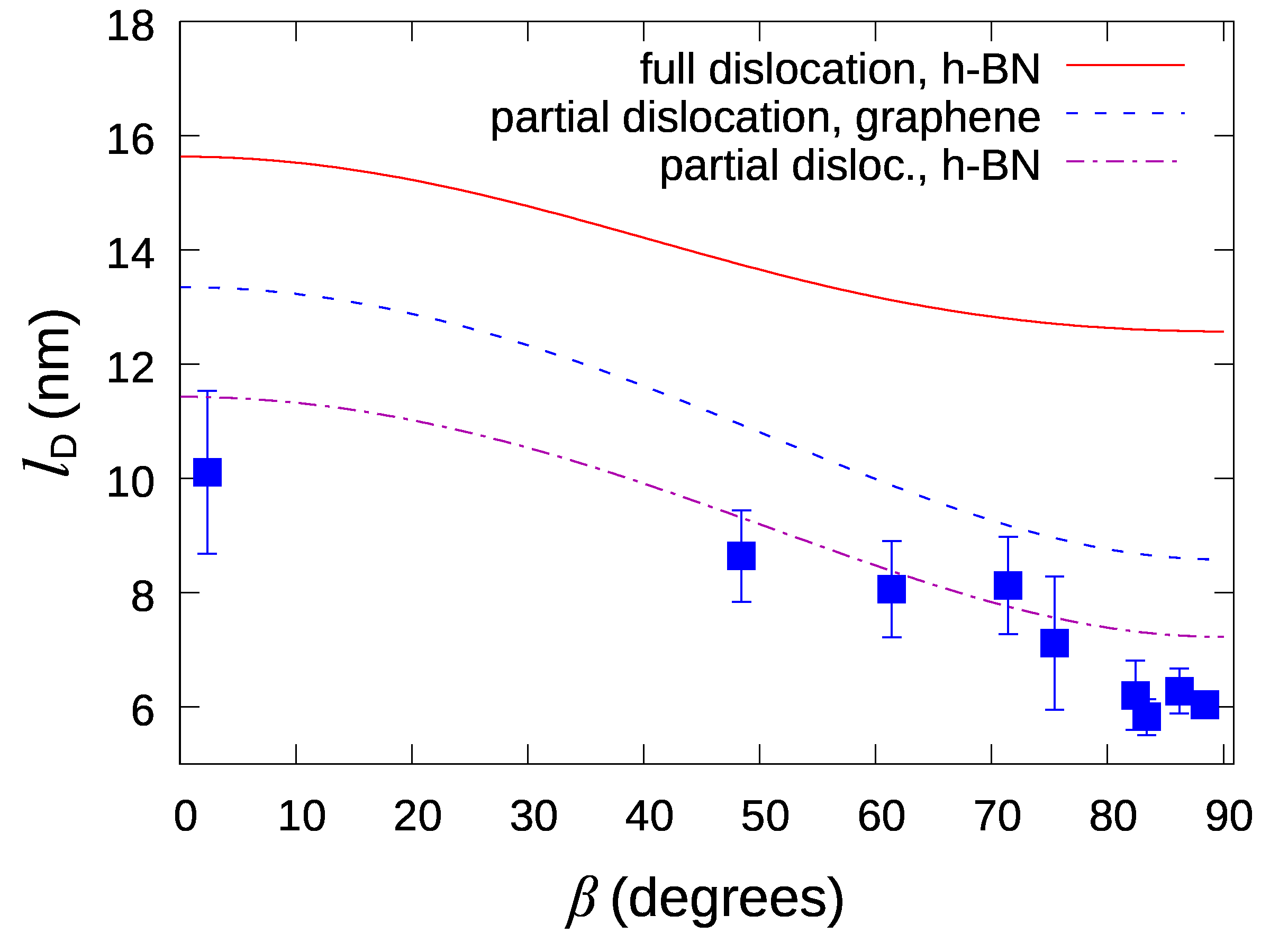}
	\caption{(Color online) Calculated dislocation width $l_\mathrm{D}$ (in nm) as a function of angle $\beta$ (in degrees)   between the Burgers vector $\vec{b}$ and normal $\vec{n}$ to the boundary between commensurate domains (Fig.~\ref{fig:statement}) for a full dislocation (red solid line) in bilayer hexagonal boron nitride with the layers aligned in the opposite directions and partial dislocations in bilayer graphene (blue dashed line) and boron nitride with the layers aligned in the same direction (violet dash-dotted line). The experimental data for graphene from Ref.~\onlinecite{Alden2013} are shown by squares ($\blacksquare$) with error bars. }
	\label{fig:width}
\end{figure}
When the interlayer interaction energy along the dislocation path is approximated simply by the cosine function, it is straightforward to derive analytically on the basis of Eq.~(\ref{eq_3a}) (for $c_1 \to 0$) that partial dislocations are described by a soliton with a short incommensurate region separating commensurate domains \cite{Chaikin1995, Popov2011}
\begin{equation} \label{eq_5}
\begin{split}
&u_x = \frac{2}{\pi}\cos{\beta}\arctan{\left[\exp{\left(2\pi x \sqrt{\frac{V_\mathrm{max}}{K(\beta)l^2}} \right)}\right]}, \\
&u_y = u_x\tan{\beta}
\end{split}
\end{equation}

Though in the present paper we describe the interlayer interaction energy in partial dislocations in a different way (Eq.~(\ref{eq_a1})) based on the approximation of the full potential energy surface of graphene and boron nitride layers aligned in the same direction, the calculated dislocation path is still very close to this analytical solution (Fig.~\ref{fig:disl_struct}d). The slope of the dependence of displacement $u_x(x)$ is nearly constant around the dislocation center and roughly equal to the maximum value $|\vec{u}'|_\mathrm{max}$. In this case the dislocation width can be defined in the way similar to the analytical solution
\begin{equation} \label{eq_5a}
	\begin{split}
		l_\mathrm{D} (\beta) = \frac{l}{\left|\vec{u}' \right|_\mathrm{max}} =\frac{l}{2} \sqrt{\frac{K(\beta)}{V_\mathrm{max}}}.
	\end{split}
\end{equation}

It is seen from this formula that the dislocation width follows that same dependence on the angle $\beta$ between the Burgers vector and normal to the boundary between commensurate domains as the formation energy $\Delta  W_0$ per unit length of the boundary between commensurate domain  (Fig.~\ref{fig:width}), increasing upon changing the dislocation character from shear to tensile. The calculated angular dependence of the dislocation width in graphene is in good agreement with the experimental data \cite{Alden2013,Lin2013,Yankowitz2014} ranging from 11 nm for tensile dislocations to 6 -- 7 nm for shear dislocations (Fig.~\ref{fig:width}). The obtained widths of partial dislocations in graphene are also on the order of the values calculated for graphite \cite{Yang2011} (5.6 nm and 3.7 nm for tensile and shear dislocations, respectively). 

For full dislocations (Fig.~\ref{fig:disl_struct}a) in boron nitride with the layers aligned in the opposite directions, we take into account contributions from two straight pieces of the dislocation path (Fig.~\ref{fig:pes}c) substituting $\sqrt{K(\beta)}$ in the expressions for the formation energy and dislocation width (Eq.~(\ref{eq_5a})) by $\sqrt{K(\beta-30^{\circ})}+\sqrt{K(\beta+ 30^{\circ})}$. The same as for partial dislocations, the formation energy per unit length of the boundary between commensurate domains is minimal for shear dislocations $W_0 ((7-\nu)/2(1-\nu^2))^{1/2}$ and maximal $W_0 ((5-3\nu)/2(1-\nu^2))^{1/2}$ for tensile dislocations (Fig.~\ref{fig:energy}), where $W_0=0.166$~eV/\AA. However, in this case the full tensile and shear dislocations correspond to the boundaries between commensurate domains in the armchair and zigzag directions, respectively. 

Similar to partial dislocations, the angular dependence of the width of full dislocations is the same as of the formation energy per unit length of the boundary between commensurate domains (Fig.~\ref{fig:width}). For each straight piece of the dislocation path, the slope $|\vec{u}' (x)|$ is nearly constant at distances from the dislocation center comparable to the dislocation width (Fig.~\ref{fig:disl_struct}c). This makes possible accurate measurements of the dislocation width. Such an information can be used to estimate the barrier to relative sliding of boron nitride layers by transmission electron microscopy in the same way as it was done for graphene \cite{Alden2013}.

\subsection{Critical elongation for formation of dislocations}
Let us consider the case when the bottom layer is stretched up to the relative elongation $\epsilon$ at an angle $\phi$ to the normal to the boundary between commensurate domains, the upper layer is free and both of the layers are allowed to relax freely in the direction perpendicular to the elongation, i.e. there are strains $\epsilon_{x'}=\epsilon$, $\epsilon_{y'}=-\nu \epsilon$ and $\gamma_{x',y'}=0$ in axes $x'$ and $y'$ along the direction of elongation and in the perpendicular direction, respectively (Fig.~\ref{fig:statement}). In axes $x$ and $y$ associated with the boundary between commensurate domains, the external strains can be written as
\begin{equation} \label{eq_6}
\begin{split}
& \epsilon_x + \nu \epsilon_y = (1 - \nu^2) \epsilon \cos^2{\phi}, \\
& \gamma_{xy} = (1+\nu) \epsilon \sin{2\phi}.
\end{split}
\end{equation}

In the case when the normal to the boundary between commensurate domains forms the angle $\beta$ with the Burgers vector $\vec{b}$ of the dislocation, the angle between the direction of elongation and Burgers vector is $\alpha = \beta - \phi$ (Fig.~\ref{fig:statement}). Using Eq.~(\ref{eq_6}) and taking into account that $\Delta u_{x} = -b/l \cos{\beta}$ and $\Delta u_{y} = -b/l \sin{\beta}$, the correction to the formation energy of such dislocations under the external strain given by Eq.~(\ref{eq_2a}) takes the form
\begin{equation} \label{eq_7}
\begin{split}
\Delta W_\epsilon (\beta, \alpha)= -\epsilon kb \cos{\alpha} \cos{(\alpha - \beta)}.
\end{split}
\end{equation}
It should be noted that for a given direction of elongation of the bottom layer, the angle $\alpha$ can take only six discrete values in two-dimensional hexagonal layers, since there are six possible directions of the Burgers vector corresponding to the armchair direction for partial dislocations or zigzag direction for full dislocations. 

\begin{figure}
	\centering
	\includegraphics[width=\columnwidth]{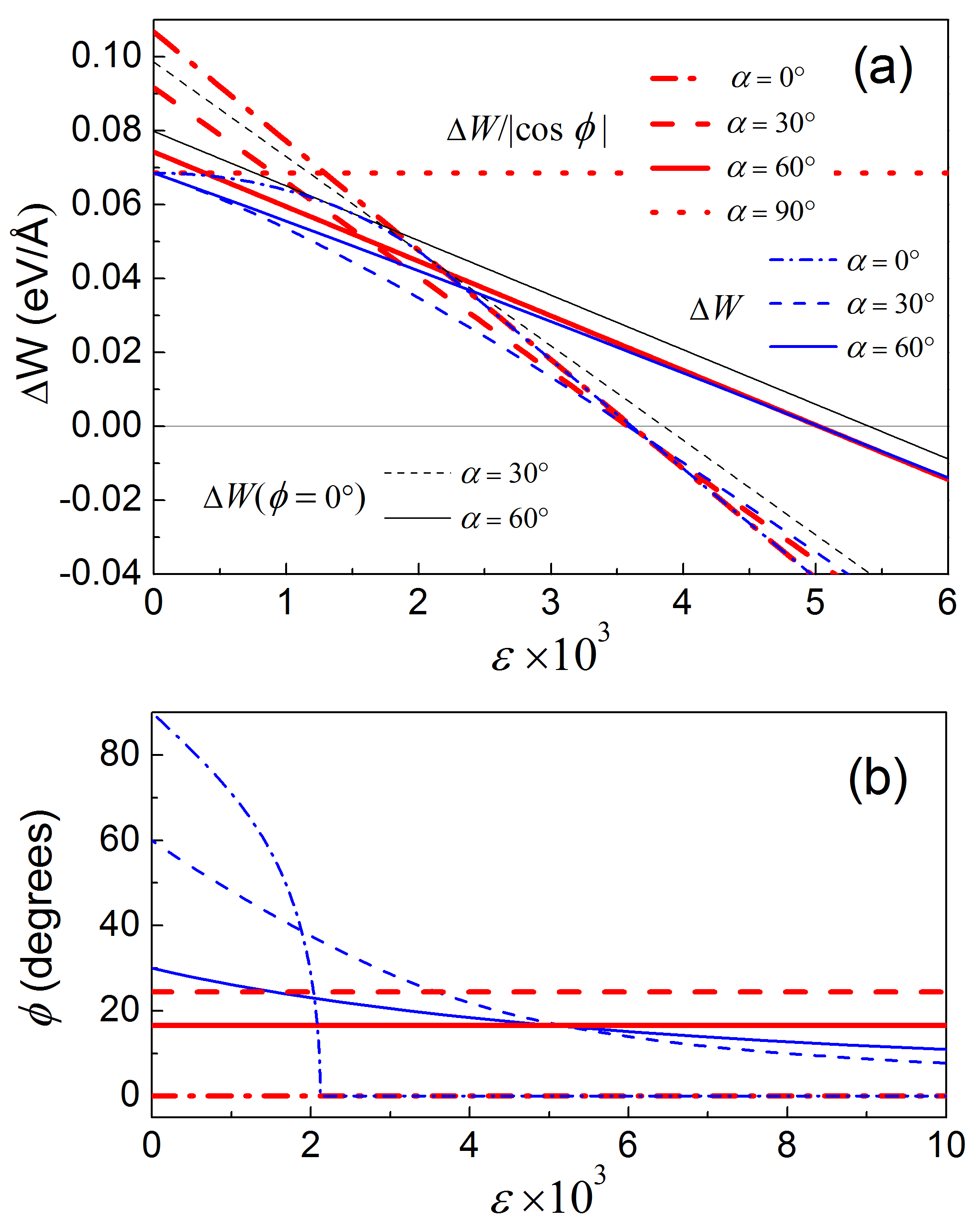}
	\caption{(Color online) (a) Formation energy $\Delta W$ (in eV/\AA) and (b) angle $\phi$ (in degrees) between the normal to the boundary between commensurate domains and direction in which the bottom layer is stretched as functions of elongation $\epsilon$ for partial dislocations in bilayer graphene. The data for the Burgers vector $\vec{b}$ at different angles $\alpha$ to the direction of elongation (Fig.~\ref{fig:statement}) are plotted: $0^{\circ}$ (dash-dotted lines), $30^{\circ}$ (dashed lines), $60^{\circ}$ (solid lines) and $90^{\circ}$ (dotted line). The thin black lines present the results for the boundaries that are perpendicular to the direction of elongation ($\phi=0^{\circ}$). The blue lines of medium thickness show the results for the boundaries between commensurate domains characterized by the minimal formation energy per unit length of the boundary. The thick red lines correspond to the results for ribbons strectched along the axis. In this case the formation energy per unit ribbon width $\Delta W/|\cos{\phi }|$ is minimized and is presented in the figure. For $\alpha=90^{\circ}$ (dotted line), all three cases correspond to the same boundary perpendicular to the direction of elongation ($\phi = 0^{\circ}$). For $\alpha=0^{\circ}$ (dash-dotted line), the minimal formation energy per ribbon width is reached for the boundary perpendicular to the direction of elongation and the thin black lines and thick red lines are the same. In these cases of coinciding lines, red lines are shown. } 
	\label{fig:energy_partial}
\end{figure}

Depending on the angle $\alpha$ between the direction of elongation and Burgers vector and elongation $\epsilon$ of the bottom layer, the optimal angles $\beta$ and $\phi = \beta - \alpha$ corresponding to the orientation of the boundary between commensurate domains (Fig.~\ref{fig:statement}) can be found by minimization of the formation energy of dislocations determined by Eqs.~(\ref{eq_4}) and (\ref{eq_7}). The dependences of the optimized formation energy $\Delta W$  of dislocations per unit length of the boundary between commensurate domains and the corresponding angle $\phi$ on the elongation $\epsilon$  
for several angles $\alpha$ are shown in Fig.~\ref{fig:energy_partial} for partial dislocations in graphene. The results for partial and full dislocations in boron nitride with the layers aligned in the same and opposite directions, respectively, are qualitatively similar. 

Except for the angle $\alpha = 90^{\circ}$, these dependences are non-linear (Fig.~\ref{fig:energy_partial}a). In the case of $\alpha = 90^{\circ}$, i.e. when the Burgers vecor is perpendicular to the direction of elongation, the formation energy cannot be reduced by stretching the bottom layer and shear dislocations with the boundary between commensurate domains parallel to the Burgers vector and perpendicular to the direction of elongation are preferred. At other angles $\alpha$, the angle $\phi$ changes from $90^{\circ}-\alpha$ in the limit of small elongations to $0^{\circ}$ in the limit of large elongations (Fig.~\ref{fig:energy_partial}b). The first of these limits means that the boundary tends to become parallel to the Burgers vector and is related to the preference of shear dislocations in the absence of external strains. In the second of these limits, the boundary between commensurate domains tends to become perpendicular to the direction of elongation. In the particular case of $\alpha = 0^{\circ}$,  i.e. when the Burgers vecor is parallel to the direction of elongation, the angle $\phi$ reaches $0^{\circ}$ at the finite elongation of $\epsilon = 0.596W_0/kl=2.12\cdot 10^{-3}$ for graphene. Upon exceeding this elongation, the boundary between commensurate domains stays perpendicular to the direction of elongation, while the formation energy depends on the elongation linearly.

The gradual reduction of the formation energy of dislocations upon increasing the elongation of the bottom layer (Eq.~(\ref{eq_7})) finally provides that the formation of dislocations becomes energetically favourable (Fig.~\ref{fig:energy_partial}a). As follows from Eqs.~(\ref{eq_4}) and (\ref{eq_7}), the critical elongation $\epsilon_\mathrm{c} (\alpha)$ at which the formation energy of dislocations with the Burgers vector at the angle $\alpha$ to the direction of elongation is zero $\Delta W = \Delta W_0+\Delta W_\epsilon=0$ is given by 
\begin{equation} \label{eq_10}
\begin{split}
\epsilon_\mathrm{c} (\alpha)= \frac{\Delta W_0 (\beta_\mathrm{c})}{kb\cos{\alpha} \cos{(\alpha - \beta_c)}}.
\end{split}
\end{equation}
Here $\beta_\mathrm{c}$ is the optimal angle between the normal to the boundary between commensurate domains and the Burgers vector of the dislocation at the critical elongation. This angle is determined by the conditions 
\begin{equation} \label{eq_8}
\begin{split}
\tan{(\alpha - \beta_\mathrm{c})} = -\frac{k\sin{2\beta_\mathrm{c}}}{4(1-\nu)K(\beta_\mathrm{c})}
\end{split}
\end{equation}
for partial dislocations (Fig.~\ref{fig:disl_struct}b) and
\begin{equation} \label{eq_9}
\begin{split}
\tan{(\alpha - \beta_\mathrm{c})} = -\frac{k\left(\sin{(2\beta_\mathrm{c}-60^{\circ})}+\kappa(\beta_\mathrm{c})\sin{(2\beta_\mathrm{c}+60^{\circ})}\right)}{4(1-\nu)K(\beta_\mathrm{c}-30^{\circ})\left(1+\kappa^{-1}(\beta_\mathrm{c})\right)}
\end{split}
\end{equation}
for full dislocations (Fig.~\ref{fig:disl_struct}a), where we use the notation $\kappa(\beta)=\sqrt{K(\beta-30^{\circ})/K(\beta+30^{\circ})}$. 

The orientation of the boundaries between commensurate domains in real samples, however, is determined by the total formation energy of dislocations, not by the energy per unit length, and depends on the sample geometry. However, the critical elongation $\epsilon_\mathrm{c}$ and the corresponding orientation $\beta_\mathrm{c}$ of the boundary between commensurate domains are always given by Eqs.~(\ref{eq_10}) -- (\ref{eq_9}), since in this case the formation energy is zero irrespective of the boundary length. 

In the particular case of a rectangular ribbon stretched along the axis (Fig.~\ref{fig:statement}), the formation energy of dislocations per unit ribbon width $\Delta W/|\cos{\phi }|=\Delta W/|\cos{(\alpha - \beta)}|$ (Fig.~\ref{fig:energy_partial}a) should be optimized. For this geometry of the sample, the optimal angles $\beta$ and $\phi = \beta - \alpha$ do not depend on the elongation (Fig.~\ref{fig:energy_partial}b) and are the same as the values given by Eqs.~(\ref{eq_8}) and (\ref{eq_9}) at the critical elongation. This leads to the linear dependence of the optimized formation energy $\Delta W/|\cos{\phi }|$ of dislocations per unit ribbon width on the elongation of the bottom layer, different from the case of the formation energy $\Delta W$ of dislocations per unit length of the boundary between commensurate domains (Fig.~\ref{fig:energy_partial}a).

\begin{figure}
	\centering
	\includegraphics[width=\columnwidth]{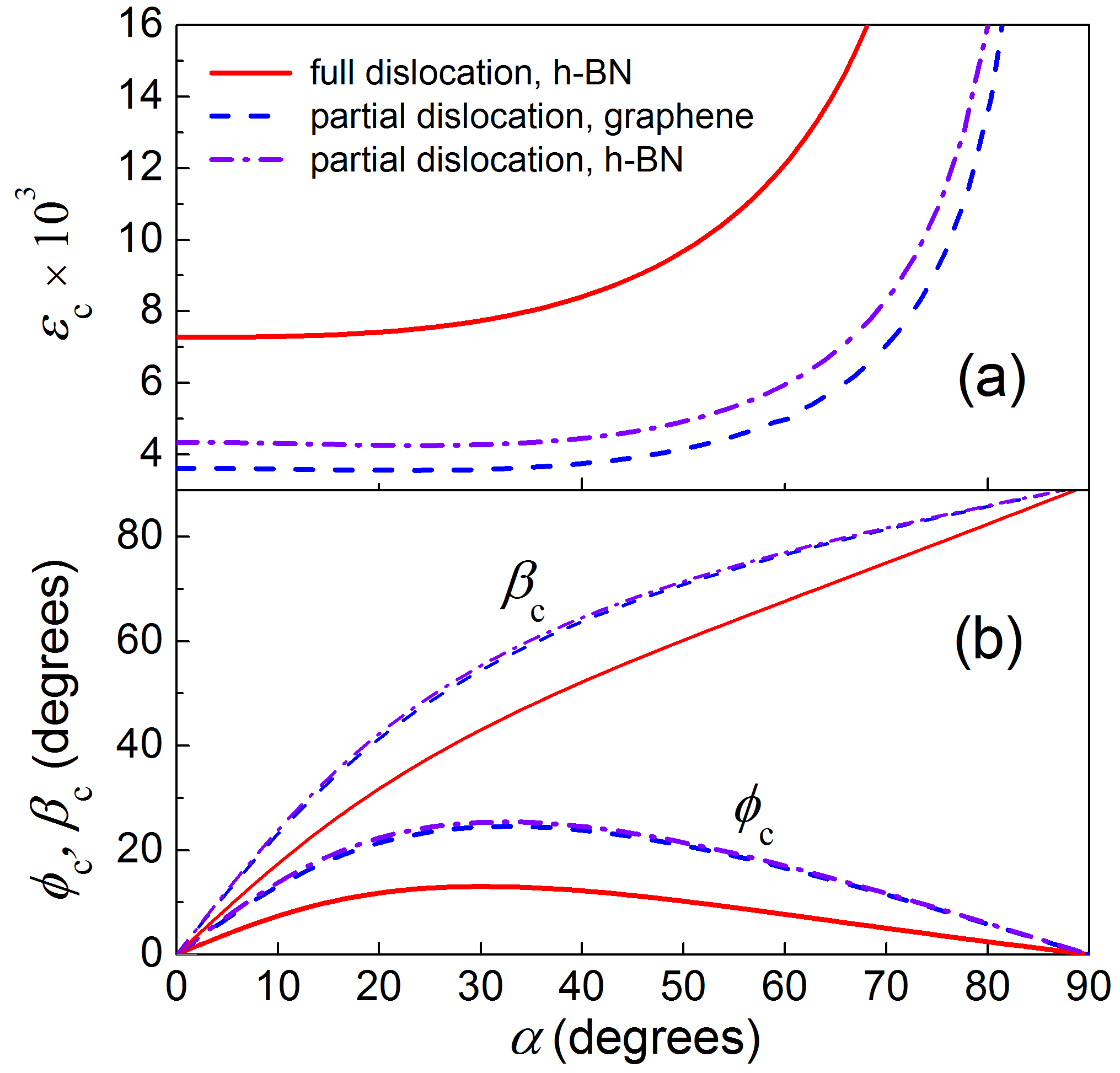}
	\caption{(Color online) (a) Critical elongation $\epsilon_\mathrm{c}$ of the bottom layer for formation of partial dislocations in bilayer graphene (blue dashed lines) and hexagonal boron nitride with the layers aligned in the same direction (violet dash-dotted lines) and full dislocations in bilayer boron nitride with the layers aligned in the opposite directions (red solid lines) as a function of the angle $\alpha$ (in degrees) between the Burgers vector and direction of elongation of the bottom layer (Fig.~\ref{fig:statement}). (b)  Optimal angles $\phi_\mathrm{c}$ between the normal to the boundary between commensurate domains and direction of elongation (in degrees, thick lines) and $\beta_\mathrm{c}$ between the normal to the boundary between commensurate domains and Burgers vector (in degrees, thin lines) at the critical elongation as functions of the angle $\alpha$ (in degrees) between the Burgers vector and direction of elongation. }
	\label{fig:critical_strain}
\end{figure}

Let us now discuss the dependence of the critical elongation $\epsilon_\mathrm{c}$ at which formation of dislocations becomes thermodynamically favourable on the angle $\alpha$  between the Burgers vector and the direction of elongation  (Fig.~\ref{fig:critical_strain}a). For all the considered materials, at $\alpha = 0^{\circ}$ this elongation is only slightly above $W_0/kl$, which corresponds to the analytical solution for partial dislocations when the interlayer interaction energy along the dislocation path is approximated by the cosine function \cite{Pokrovsky1978,Popov2011}. In the limit of $\alpha \to 90^{\circ}$, the critical elongation becomes infinitely large, demonstrating that it is not possible to reduce the formation energy of dislocations by stretching the bottom layer in the direction orthogonal to the Burgers vector. The optimal angle $\phi_\mathrm{c}$ between the boundary between commensurate domains and the direction of elongation at the critical elongation is 0 in the limits $\alpha \to 0^{\circ}$ and $\alpha \to 90^{\circ}$ (Fig.~\ref{fig:critical_strain}b), i.e. the boundary between commensurate domains tends to be perpendicular to the direction of elongation in these limits. 

In spite of these qualitative similarities in the dependences of the critical elongation and orientation of the boundary between commensurate domains on the direction of elongation for the considered materials, they are some differences. The critical elongation for full dislocations in bilayer hexagonal boron nitride with the layers aligned in the opposite directions exceeds  $W_0/kl=6.69 \cdot 10^{-3}$ by 9\% at $\alpha = 0^{\circ}$ and grows monotonically. For partial dislocations in bilayer graphene or boron nitride with the layers aligned in the same direction, the critical elongation is only 2\% greater than $W_0/kl$ at $\alpha = 0^{\circ}$ and weakly depends on $\alpha$ for $\alpha<60^{\circ}$. Nevertheless, there is a small minimum in the critical elongation corresponding exactly to the value $W_0/kl = 3.56 \cdot 10^{-3}$ at $\alpha=22.6^{\circ}$ for bilayer graphene. For boron nitride with the layers aligned in the same direction, this minimum is $W_0/kl = 4.25\cdot 10^{-3}$ and it is observed at $\alpha=24.1^{\circ}$. Smaller values of the critical elongation for graphene compared to boron nitride with the layers aligned in the same direction is related to the higher stiffness of graphene. 

\subsection{Phase diagram}
\begin{figure}
	\centering
	\includegraphics[width=\columnwidth]{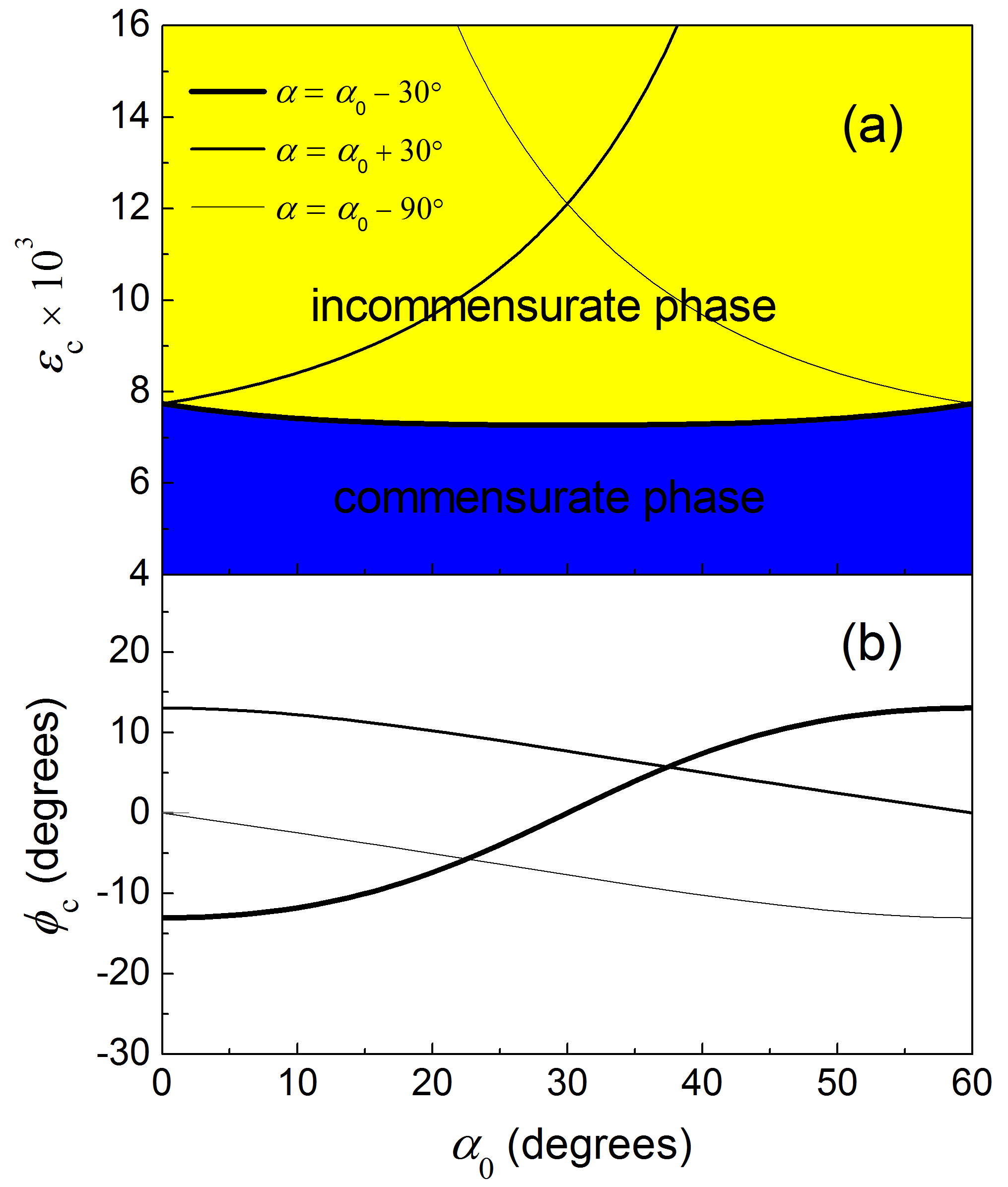}
	\caption{(Color online) (a) Critical elongation $\epsilon_\mathrm{c}$ of the bottom layer for formation of full dislocations in bilayer hexagonal boron nitride with the layers aligned in the opposite directions and (b) optimal angle $\phi_\mathrm{c}$ (in degrees) between the normal to the boundary between commensurate domains and direction in which the bottom layer is stretched  (Fig.~\ref{fig:statement}) at the critical elongation as functions of the angle $\alpha_0$ (in degrees) between the direction of elongation and any of the armchair directions: (thick lines) $\alpha = \alpha_0 - 30^{\circ}$, (medium lines)   $\alpha = \alpha_0 + 30^{\circ}$ and (thin lines) $\alpha = \alpha_0 - 90^{\circ}$. The following phases are indicated: (blue/dark gray) commensurate phase and (yellow/light gray) incommensurate phase with multiple full dislocations with the angle $\alpha = \alpha_0 - 30^{\circ}$. }
	\label{fig:critical_strain_full}
\end{figure}

Let us now consider different phases that are possible in two-dimensional hexagonal bilayers. For boron nitride bilayer with the layers aligned in the opposite directions, where elementary dislocations are full, the phase diagram is rather simple since adjacent dislocations can have the same Burgers vector. Then to minimize the total energy all dislocations present in the sample should be identical. Due to the symmetry of the layers it makes sense to consider only the angles $0^{\circ} \le \alpha_0 \le 30^{\circ}$ between the direction of elongation and any of the armchair directions. At each angle $\alpha_0$ there are three possible full dislocations with the positive critical elongation (Fig.~\ref{fig:critical_strain_full}a, the diagram for the negative elongations is similar).
 The smallest critical elongation $\tilde{\epsilon}_\mathrm{c}(\alpha_0)$ corresponding to the dislocations with the angle $\alpha = \alpha_0 - 30^{\circ}$ is the one at which formation of the first dislocation takes place (Fig.~\ref{fig:critical_strain_full}b). Increasing the elongation of the bottom layer further makes favourable incorporation of more and more similar dislocations \cite{Chaikin1995}. Therefore, the formation of the first dislocation in bilayer boron nitride bilayer with the layers aligned in the opposite directions corresponds to the second-order phase transition to the incommensurate phase described by the density of dislocations as the order parameter \cite{Pokrovsky1978,Popov2011}. The optimal structures at elongations exceeding $\tilde{\epsilon}_\mathrm{c}(\alpha_0)$ contain multiple identical dislocations characterized by the same angle $\alpha = \alpha_0 - 30^{\circ}$ and the same orientation of the boundary between commensurate domains, i.e. angles $\beta(\alpha, \epsilon)$ and $\phi(\alpha, \epsilon)$ (Fig.~\ref{fig:statement}). The critical elongation  $\tilde{\epsilon}_\mathrm{c}$ weakly depends on the direction of elongation and is about $7.5 \cdot 10^{-3}$. A shallow minimum in the critical elongation is reached for the bottom layer stretched in the zigzag direction.

\begin{figure}
	\centering
	\includegraphics[width=\columnwidth]{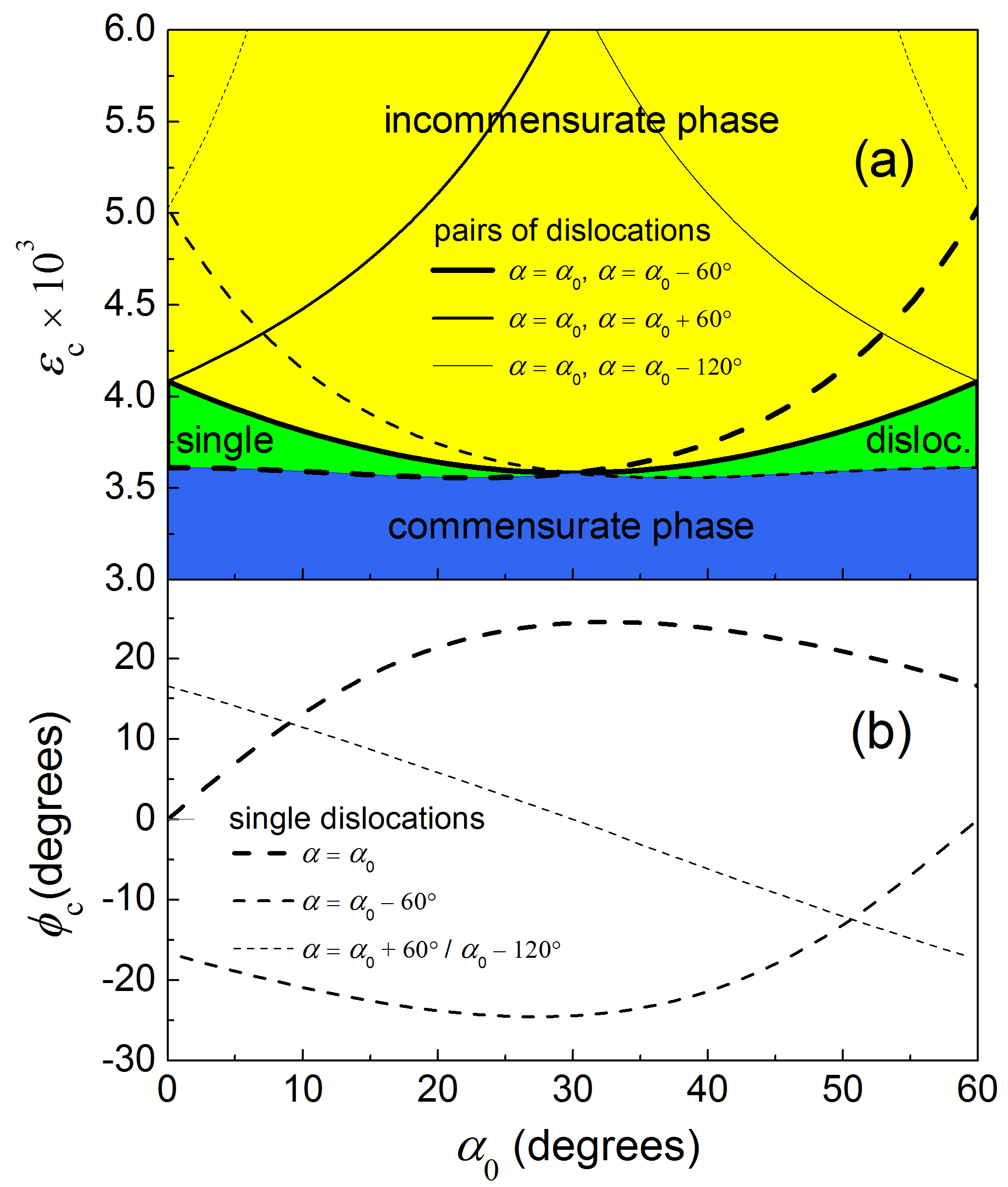}
	\caption{(Color online) (a) Critical elongation $\epsilon_\mathrm{c}$ of the bottom layer for formation of single partial dislocations (dashed lines) and pairs of partial dislocations (solid lines) in bilayer graphene as a function of the angle $\alpha_0$ (in degrees) between the direction of elongation and any of the armchair directions. Single partial dislocations: (thick dashed lines) $\alpha = \alpha_0$, (medium dashed lines) $\alpha = \alpha_0 - 60^{\circ}$, (thin dashed lines) $\alpha = \alpha_0 + 60^{\circ}$ (for $\alpha_0 < 30^{\circ}$) and $\alpha = \alpha_0 - 120^{\circ}$ (for $\alpha_0 > 30^{\circ}$). Pairs of partial dislocations include one dislocation with $\alpha = \alpha_0$ and another dislocation with (thick solid line) $\alpha = \alpha_0 - 60^{\circ}$, (medium solid line) $\alpha = \alpha_0 + 60^{\circ}$ or (thin solid line) $\alpha = \alpha_0 - 120^{\circ}$. (b) Optimal angle $\phi_\mathrm{c}$  (in degrees) between the normal to the boundary between commensurate domains and direction of elongation (Fig.~\ref{fig:statement}) at the critical elongation for single partial dislocations. The following phases are indicated: (blue/dark gray) commensurate phase, (green/medium gray) structure with a single partial dislocation with $\alpha = \alpha_0$ and (yellow/light gray) incommensurate phase with multiple pairs of partial dislocations with $\alpha = \alpha_0$ and $\alpha = \alpha_0 - 60^{\circ}$. The results for pairs of partial dislocations are presented for bilayers of the ribbon shape. }
	\label{fig:critical_strain_arm}
\end{figure}

For graphene and boron nitride with the layers aligned in the same direction, there are again three possible partial dislocations with different directions of the Burgers vector at each angle $\alpha_0$ (Fig.~\ref{fig:critical_strain_arm}a). Formation of the first dislocation takes place at the smallest of the corresponding critical elongations $\tilde{\epsilon}_\mathrm{c}(\alpha_0)$ reached for dislocations with $\alpha = \alpha_0$ (Fig.~\ref{fig:critical_strain_arm}b). However, generation of adjacent identical partial dislocations is not possible. The potential surface of interlayer energy (Fig.~\ref{fig:pes}a) requires that Burgers vectors of consecutive partial dislocation change so that their sum gives the vector with the magnitude of the lattice constant. Therefore, transition to the incommensurate phase with multiple partial dislocations occurs at a higher elongation (Fig.~\ref{fig:critical_strain_arm}a) when penetration of pairs of partial dislocations with the sum of the Burgers vector equal in magnitude to the lattice constant becomes thermodynamically favourable. In the case of bilayer samples of the ribbon shape, this second critical elongation is determined by the equation similar to Eq.~(\ref{eq_10}), where the formation energy $\Delta W_0/|\cos{\phi}|=\Delta W_0/|\cos{(\alpha - \beta)}|$ of dislocations per unit ribbon width and the projection of the Burgers vectors on the ribbon axis $b\cos{\alpha}$ are substituted by the sum of the corresponding quantities for the pair of dislocations. The optimal pair of partial dislocations in this case is characterized by the angles $\alpha = \alpha_0$ and $\alpha = \alpha_0 - 60^{\circ}$. The requirement of alteration of the Burgers vectors of partial dislocations in the incommensurate phase of graphene and boron nitride with layers aligned in the same direction also leads to alterating orientations of the boundaries between commensurate domains (Fig.~\ref{fig:statement}).

It should be noted that the first critical elongation $\tilde{\epsilon}_\mathrm{c}$ for partial dislocations weakly depends on the direction of elongation of the bottom layer (Fig.~\ref{fig:critical_strain_arm}a) and is about $3.6 \cdot 10^{-3}$ and $4.3 \cdot 10^{-3}$ for bilayer graphene and boron nitride with the layers aligned in the same direction, respectively. The second critical elongation for generation of multiple partial dislocations has a clear minimum when the bottom layer is stretched the zigzag direction. In this case two different partial dislocations with the Burgers vectors at angles $\alpha = \pm 30^{\circ}$ to the direction of elongation are equal in energy and become thermodynamically favourable at the same elongation. Therefore, when the bottom layer is stretched in the zigzag direction, the commensurate-incommensurate phase transition occurs at the same time as the formation of the first dislocation, similar to boron nitride with the layers aligned in the opposite directions.

\section{Conclusions}
We have used potential energy surfaces of bilayer graphene and hexagonal boron nitride obtained by DFT calculations to analyze formation of dislocations in stacking of the layers. The two-chain Frenkel-Kontorova model has been extended to describe dislocations with arbitrary relative orientations of the Burgers vector, boundary between commensurate domains and external strain. The dislocation width, orientation of the boundary between commensurate domains and structural transformations upon increasing the elongation of the bottom layer have been studied. 

It is suggested that along with graphene, partial dislocations are possible in metastable boron nitride with the layers aligned in the same direction (AB stacking in the commensurate state). The characteristic width and formation energy of such dislocations at zero external strain are found to be rather close for these two similar materials. The calculated width of partial dislocations in graphene is in good agreement with the experimental data.

In bilayer boron nitride with the layers aligned in opposite directions, formation of the first full dislocation and commensurate-incommensurate phase transition are found to occur at the same critical elongation. Above this critical elongation, similar dislocations with the same Burgers vector and orientation of the boundary between commensurate domains are expected to form. It is revealed that the critical elongation weakly depends on the direction of elongation. However, there is a shallow minimum for the elongation in the zigzag direction.

In the case of bilayer graphene or boron nitride with the layers aligned in the same direction, the second-order transition to the incommensurate phase with the density of dislocations changing continously with the elongation is preceded by the transition to the state with only one partial dislocation. Partial dislocations in the incommensurate phase are characterized by alterating orientations of the Burgers vector and, consequently, alterating orientations of the boundaries between commensurate domains. The first critical elongation corresponding to the formation of a single partial dislocation shows a weak dependence on the direction of elongation. The second critical elongation at which the transition to the incommensurate phase with multiple partial dislocations occurs has a clear minimum for the elongation in the zigzag direction. Bilayer graphene is found to show slightly greater critical elongations compared to boron nitride with the layers aligned in the same direction due to the higher stiffness of graphene.

Electronic properties of monolayer graphene deposited on a flexible substrate at uniaxial strains up to 0.8\% have been studied by stretching the substrate in one direction \cite{Ni2008}. A method of production of bilayer graphene nanoribbons by deposition on substrates has been elaborated \cite{Li2008}. The boundaries between commensurate domains of bilayer and few-layer graphene have been investigated by scanning transmission microscopy \cite{Alden2013,Lin2013} and scanning tunneling microscopy \cite{Yankowitz2014}. These advances in the experimental techniques give us a cause for the optimism that the considered commensurate-incommensurate transition can be observed experimentally for two-dimensional bilayer crystals of the ribbon shape.

\begin{acknowledgements}
The authors gratefully thank Prof. Yu. E. Lozovik for valuable discussions. The authors also acknowledge the Russian Foundation of Basic Research (14-02-00739-a) and computational time on the Supercomputing Center of Lomonosov Moscow State University and the Multipurpose Computing Complex NRC ``Kurchatov Institute". IL acknowledges the financial support from Grupos Consolidados UPV/EHU del Gobierno Vasco (IT578-13) and H2020-NMP-2014 project ``MOSTOPHOS" (n. 646259). 
\end{acknowledgements}

\appendix*
\section{Dislocation path}
The structure of the dislocations, i.e. the dependences of the relative $\vec{u}$ and averaged $\vec{a}$ displacements of the layers on the coordinate $x$ in the direction perpendicular to the boundary between commensurate domains that minimize the formation energy of dislocations (Eqs.~(\ref{eq_2}) and (\ref{eq_2a})), are determined by the Euler-Lagrange equations 
\begin{equation} \label{eq_3}
\begin{split}
&\frac{\partial \Delta W}{\partial u_x}=-\frac{1}{2} E u''_x+ \frac{\partial V(\vec{u})}{\partial u_x}=0, \\
&\frac{\partial \Delta W}{\partial u_y}=-\frac{1}{2} G u''_y+ \frac{\partial V(\vec{u})}{\partial u_y}=0, \\
&\frac{\partial \Delta W}{\partial a_x}=-\frac{1}{2} E a''_x =0, \quad \frac{\partial \Delta W}{\partial a_y}=-\frac{1}{2} G a''_y=0.
\end{split}
\end{equation}
The external strain applied to the bottom layer does not enter in these equations. Therefore, the dislocation structure under external strain is not different from the one in the absence of the strain. 

The first two of the Euler-Lagrange equations~(\ref{eq_3}) show that the dislocation path is the same as the trajectory of a particle on the inverse potential energy surface $-V(\vec{u})$ with the fixed initial and final positions corresponding to the deepest energy minima of the original potential energy surface. It should be also noted, however, that the ``particle mass" is anisotropic and equal to $G/2$ and $E/2$ in the directions along and across the boundary between commensurate domains, respectively. In the case of graphene or boron nitride layers aligned in the same direction this means that the dislocation path is described exactly by the straight line between the adjacent AB minima (Fig.~\ref{fig:pes}a), i.e. the minimum energy path \cite{Lebedev2015, Popov2011}. 

The solution of Eq.~(\ref{eq_3}) for full dislocations in boron nitride layers aligned in the opposite directions, on the other hand, depends on the angle between the Burgers vector and the boundary between commensurate domains and cannot be derived analytically. Based on the qualitative considerations that the rapidly changing potential at slopes of the AB2' hill should be avoided, we assume that the path of full dislocations is also roughly described by the minimum energy path AA' -- AB1' -- AA' consisting of two straight lines at the angle $120^{\circ}$ to each other (Fig.~\ref{fig:pes}c). A more accurate approximation of the dislocation path can be proposed by introduction of a shortcut between AA' -- AB1' and AB1' -- AA' lines avoiding the local energy minimum AB1'. The position of this additional straight piece is optimized to minimize the formation energy of dislocations in the absence of external strains. However, the correction to the formation energy in this case does not exceed 10\%. This provides an estimate of the accuracy of our approximation of the dislocation path as AA' -- AB1' -- AA'. It should be noted that this accuracy is comparable to that of DFT calculations of the potential surface of interlayer interaction energy, which can be deduced by comparison of the physically measurable quantities, such as shear mode frequency \cite{Popov2012,Lebedev2015}, dislocation width of partial dislocations \cite{Alden2013} and shear modulus \cite{Lebedev2015}, with the experimental data. 

\bibliography{rsc}

\begin{thebibliography}{62}%
\makeatletter
\providecommand \@ifxundefined [1]{%
 \@ifx{#1\undefined}
}%
\providecommand \@ifnum [1]{%
 \ifnum #1\expandafter \@firstoftwo
 \else \expandafter \@secondoftwo
 \fi
}%
\providecommand \@ifx [1]{%
 \ifx #1\expandafter \@firstoftwo
 \else \expandafter \@secondoftwo
 \fi
}%
\providecommand \natexlab [1]{#1}%
\providecommand \enquote  [1]{``#1''}%
\providecommand \bibnamefont  [1]{#1}%
\providecommand \bibfnamefont [1]{#1}%
\providecommand \citenamefont [1]{#1}%
\providecommand \href@noop [0]{\@secondoftwo}%
\providecommand \href [0]{\begingroup \@sanitize@url \@href}%
\providecommand \@href[1]{\@@startlink{#1}\@@href}%
\providecommand \@@href[1]{\endgroup#1\@@endlink}%
\providecommand \@sanitize@url [0]{\catcode `\\12\catcode `\$12\catcode
  `\&12\catcode `\#12\catcode `\^12\catcode `\_12\catcode `\%12\relax}%
\providecommand \@@startlink[1]{}%
\providecommand \@@endlink[0]{}%
\providecommand \url  [0]{\begingroup\@sanitize@url \@url }%
\providecommand \@url [1]{\endgroup\@href {#1}{\urlprefix }}%
\providecommand \urlprefix  [0]{URL }%
\providecommand \Eprint [0]{\href }%
\providecommand \doibase [0]{http://dx.doi.org/}%
\providecommand \selectlanguage [0]{\@gobble}%
\providecommand \bibinfo  [0]{\@secondoftwo}%
\providecommand \bibfield  [0]{\@secondoftwo}%
\providecommand \translation [1]{[#1]}%
\providecommand \BibitemOpen [0]{}%
\providecommand \bibitemStop [0]{}%
\providecommand \bibitemNoStop [0]{.\EOS\space}%
\providecommand \EOS [0]{\spacefactor3000\relax}%
\providecommand \BibitemShut  [1]{\csname bibitem#1\endcsname}%
\let\auto@bib@innerbib\@empty
\bibitem [{\citenamefont {Chaikin}\ and\ \citenamefont
  {Lubensky}(1995)}]{Chaikin1995}%
  \BibitemOpen
  \bibfield  {author} {\bibinfo {author} {\bibfnamefont {P.~M.}\ \bibnamefont
  {Chaikin}}\ and\ \bibinfo {author} {\bibfnamefont {T.~C.}\ \bibnamefont
  {Lubensky}},\ }\href@noop {} {\emph {\bibinfo {title} {Principles of
  Condensed Matter Physics}}}\ (\bibinfo  {publisher} {Cambridge University
  Press, Cambridge},\ \bibinfo {year} {1995})\BibitemShut {NoStop}%
\bibitem [{\citenamefont {Pennycook}(2008)}]{Pennycook2008}%
  \BibitemOpen
  \bibfield  {author} {\bibinfo {author} {\bibfnamefont {S.~J.}\ \bibnamefont
  {Pennycook}},\ }\bibfield  {title} {\enquote {\bibinfo {title} {Investigating
  the optical properties of dislocations by scanning transmission electron
  microscopy},}\ }\href {\doibase 10.1002/sca.20114} {\bibfield  {journal}
  {\bibinfo  {journal} {Scanning}\ }\textbf {\bibinfo {volume} {30}},\ \bibinfo
  {pages} {287--298} (\bibinfo {year} {2008})}\BibitemShut {NoStop}%
\bibitem [{\citenamefont {Shreter}\ \emph {et~al.}(1993)\citenamefont
  {Shreter}, \citenamefont {Rebane},\ and\ \citenamefont
  {Peaker}}]{Shreter1993}%
  \BibitemOpen
  \bibfield  {author} {\bibinfo {author} {\bibfnamefont {Yu.~G.}\ \bibnamefont
  {Shreter}}, \bibinfo {author} {\bibfnamefont {Yu.~T.}\ \bibnamefont
  {Rebane}}, \ and\ \bibinfo {author} {\bibfnamefont {A.~R.}\ \bibnamefont
  {Peaker}},\ }\bibfield  {title} {\enquote {\bibinfo {title} {Optical
  properties of dislocations in silicon crystals},}\ }\href {\doibase
  10.1002/pssa.2211380239} {\bibfield  {journal} {\bibinfo  {journal} {Physica
  Status Solidi (a)}\ }\textbf {\bibinfo {volume} {138}},\ \bibinfo {pages}
  {681--686} (\bibinfo {year} {1993})}\BibitemShut {NoStop}%
\bibitem [{\citenamefont {Holt}\ and\ \citenamefont {Yacobi}(2007)}]{Holt2007}%
  \BibitemOpen
  \bibfield  {author} {\bibinfo {author} {\bibfnamefont {D.~B.}\ \bibnamefont
  {Holt}}\ and\ \bibinfo {author} {\bibfnamefont {B.~G.}\ \bibnamefont
  {Yacobi}},\ }\href@noop {} {\emph {\bibinfo {title} {Extended defects in
  semiconductors: Electronic properties, device effects and structures}}}\
  (\bibinfo  {publisher} {Cambridge University Press, Cambridge},\ \bibinfo
  {year} {2007})\BibitemShut {NoStop}%
\bibitem [{\citenamefont {Chernatynskiy}\ \emph {et~al.}(2012)\citenamefont
  {Chernatynskiy}, \citenamefont {Clarke},\ and\ \citenamefont
  {Phillpot}}]{Chernatynskiy2012}%
  \BibitemOpen
  \bibfield  {author} {\bibinfo {author} {\bibfnamefont {A.}~\bibnamefont
  {Chernatynskiy}}, \bibinfo {author} {\bibfnamefont {D.~R.}\ \bibnamefont
  {Clarke}}, \ and\ \bibinfo {author} {\bibfnamefont {S.~R.}\ \bibnamefont
  {Phillpot}},\ }\bibfield  {title} {\enquote {\bibinfo {title} {Thermal
  transport in nanostructured materials},}\ }in\ \href@noop {} {\emph {\bibinfo
  {booktitle} {Handbook of Nanoscience, Engineering, and Technology, Third
  Edition}}},\ \bibinfo {editor} {edited by\ \bibinfo {editor} {\bibfnamefont
  {William A.~Goddard}\ \bibnamefont {III}}, \bibinfo {editor} {\bibfnamefont
  {Donald~W.}\ \bibnamefont {Brenner}}, \bibinfo {editor} {\bibfnamefont
  {Sergey~E.}\ \bibnamefont {Lyshevski}}, \ and\ \bibinfo {editor}
  {\bibfnamefont {Gerald~J.}\ \bibnamefont {Iafrate}}}\ (\bibinfo  {publisher}
  {CRC Press},\ \bibinfo {year} {2012})\BibitemShut {NoStop}%
\bibitem [{\citenamefont {Alden}\ \emph {et~al.}(2013)\citenamefont {Alden},
  \citenamefont {Tsen}, \citenamefont {Huang}, \citenamefont {Hovden},
  \citenamefont {Brown}, \citenamefont {Park}, \citenamefont {Muller},\ and\
  \citenamefont {McEuen}}]{Alden2013}%
  \BibitemOpen
  \bibfield  {author} {\bibinfo {author} {\bibfnamefont {J.~S.}\ \bibnamefont
  {Alden}}, \bibinfo {author} {\bibfnamefont {A.~W.}\ \bibnamefont {Tsen}},
  \bibinfo {author} {\bibfnamefont {P.~Y.}\ \bibnamefont {Huang}}, \bibinfo
  {author} {\bibfnamefont {R.}~\bibnamefont {Hovden}}, \bibinfo {author}
  {\bibfnamefont {L.}~\bibnamefont {Brown}}, \bibinfo {author} {\bibfnamefont
  {J.}~\bibnamefont {Park}}, \bibinfo {author} {\bibfnamefont {D.~A.}\
  \bibnamefont {Muller}}, \ and\ \bibinfo {author} {\bibfnamefont {P.~L.}\
  \bibnamefont {McEuen}},\ }\bibfield  {title} {\enquote {\bibinfo {title}
  {Strain solitons and topological defects in bilayer graphene},}\ }\href
  {\doibase 10.1073/pnas.1309394110} {\bibfield  {journal} {\bibinfo  {journal}
  {PNAS}\ }\textbf {\bibinfo {volume} {110}},\ \bibinfo {pages} {11256--11260}
  (\bibinfo {year} {2013})}\BibitemShut {NoStop}%
\bibitem [{\citenamefont {Butz}\ \emph {et~al.}(2014)\citenamefont {Butz},
  \citenamefont {Dolle}, \citenamefont {Niekiel}, \citenamefont {Weber},
  \citenamefont {Waldmann}, \citenamefont {Weber}, \citenamefont {Meyer},\ and\
  \citenamefont {Spiecker}}]{Butz2014}%
  \BibitemOpen
  \bibfield  {author} {\bibinfo {author} {\bibfnamefont {B.}~\bibnamefont
  {Butz}}, \bibinfo {author} {\bibfnamefont {C.}~\bibnamefont {Dolle}},
  \bibinfo {author} {\bibfnamefont {F.}~\bibnamefont {Niekiel}}, \bibinfo
  {author} {\bibfnamefont {K.}~\bibnamefont {Weber}}, \bibinfo {author}
  {\bibfnamefont {D.}~\bibnamefont {Waldmann}}, \bibinfo {author}
  {\bibfnamefont {H.~B.}\ \bibnamefont {Weber}}, \bibinfo {author}
  {\bibfnamefont {B.}~\bibnamefont {Meyer}}, \ and\ \bibinfo {author}
  {\bibfnamefont {E.}~\bibnamefont {Spiecker}},\ }\bibfield  {title} {\enquote
  {\bibinfo {title} {Dislocations in bilayer graphene},}\ }\href {\doibase
  10.1038/nature12780} {\bibfield  {journal} {\bibinfo  {journal} {Nature}\
  }\textbf {\bibinfo {volume} {505}},\ \bibinfo {pages} {533--537} (\bibinfo
  {year} {2014})}\BibitemShut {NoStop}%
\bibitem [{\citenamefont {Lin}\ \emph {et~al.}(2013)\citenamefont {Lin},
  \citenamefont {Fang}, \citenamefont {Zhou}, \citenamefont {Lupini},
  \citenamefont {Idrobo}, \citenamefont {Kong}, \citenamefont {Pennycook},\
  and\ \citenamefont {Pantelides}}]{Lin2013}%
  \BibitemOpen
  \bibfield  {author} {\bibinfo {author} {\bibfnamefont {J.}~\bibnamefont
  {Lin}}, \bibinfo {author} {\bibfnamefont {W.}~\bibnamefont {Fang}}, \bibinfo
  {author} {\bibfnamefont {W.}~\bibnamefont {Zhou}}, \bibinfo {author}
  {\bibfnamefont {A.~R.}\ \bibnamefont {Lupini}}, \bibinfo {author}
  {\bibfnamefont {J.~C.}\ \bibnamefont {Idrobo}}, \bibinfo {author}
  {\bibfnamefont {J.}~\bibnamefont {Kong}}, \bibinfo {author} {\bibfnamefont
  {S.~J.}\ \bibnamefont {Pennycook}}, \ and\ \bibinfo {author} {\bibfnamefont
  {S.~T.}\ \bibnamefont {Pantelides}},\ }\bibfield  {title} {\enquote {\bibinfo
  {title} {{AC/AB} stacking boundaries in bilayer graphene},}\ }\href {\doibase
  10.1021/nl4013979} {\bibfield  {journal} {\bibinfo  {journal} {Nano Letters}\
  }\textbf {\bibinfo {volume} {13}},\ \bibinfo {pages} {3262--3268} (\bibinfo
  {year} {2013})}\BibitemShut {NoStop}%
\bibitem [{\citenamefont {Yankowitz}\ \emph {et~al.}(2014)\citenamefont
  {Yankowitz}, \citenamefont {Wang}, \citenamefont {Birdwell}, \citenamefont
  {Chen}, \citenamefont {Watanabe}, \citenamefont {Taniguchi}, \citenamefont
  {Jacquod}, \citenamefont {San-Jose}, \citenamefont {Jarillo-Herrero},\ and\
  \citenamefont {LeRoy}}]{Yankowitz2014}%
  \BibitemOpen
  \bibfield  {author} {\bibinfo {author} {\bibfnamefont {M.}~\bibnamefont
  {Yankowitz}}, \bibinfo {author} {\bibfnamefont {J.~I-J.}\ \bibnamefont
  {Wang}}, \bibinfo {author} {\bibfnamefont {A.~G.}\ \bibnamefont {Birdwell}},
  \bibinfo {author} {\bibfnamefont {Yu-An}\ \bibnamefont {Chen}}, \bibinfo
  {author} {\bibfnamefont {K.}~\bibnamefont {Watanabe}}, \bibinfo {author}
  {\bibfnamefont {T.}~\bibnamefont {Taniguchi}}, \bibinfo {author}
  {\bibfnamefont {P.}~\bibnamefont {Jacquod}}, \bibinfo {author} {\bibfnamefont
  {P.}~\bibnamefont {San-Jose}}, \bibinfo {author} {\bibfnamefont
  {P.}~\bibnamefont {Jarillo-Herrero}}, \ and\ \bibinfo {author} {\bibfnamefont
  {B.~J.}\ \bibnamefont {LeRoy}},\ }\bibfield  {title} {\enquote {\bibinfo
  {title} {Electric field control of soliton motion and stacking in trilayer
  graphene},}\ }\href {\doibase 10.1038/nmat3965} {\bibfield  {journal}
  {\bibinfo  {journal} {Nat. Mater.}\ }\textbf {\bibinfo {volume} {13}},\
  \bibinfo {pages} {786--789} (\bibinfo {year} {2014})}\BibitemShut {NoStop}%
\bibitem [{\citenamefont {Skowron}\ \emph {et~al.}(2015)\citenamefont
  {Skowron}, \citenamefont {Lebedeva}, \citenamefont {Popov},\ and\
  \citenamefont {Bichoutskaia}}]{Skowron2015}%
  \BibitemOpen
  \bibfield  {author} {\bibinfo {author} {\bibfnamefont {S.~T.}\ \bibnamefont
  {Skowron}}, \bibinfo {author} {\bibfnamefont {I.~V.}\ \bibnamefont
  {Lebedeva}}, \bibinfo {author} {\bibfnamefont {A.~M.}\ \bibnamefont {Popov}},
  \ and\ \bibinfo {author} {\bibfnamefont {E.}~\bibnamefont {Bichoutskaia}},\
  }\bibfield  {title} {\enquote {\bibinfo {title} {Energetics of atomic scale
  structure changes in graphene},}\ }\href {\doibase 10.1039/c4cs00499j}
  {\bibfield  {journal} {\bibinfo  {journal} {Chem. Soc. Rev.}\ }\textbf
  {\bibinfo {volume} {44}},\ \bibinfo {pages} {3134--3176} (\bibinfo {year}
  {2015})}\BibitemShut {NoStop}%
\bibitem [{\citenamefont {Popov}\ \emph {et~al.}(2011)\citenamefont {Popov},
  \citenamefont {Lebedeva}, \citenamefont {Knizhnik}, \citenamefont {Lozovik},\
  and\ \citenamefont {Potapkin}}]{Popov2011}%
  \BibitemOpen
  \bibfield  {author} {\bibinfo {author} {\bibfnamefont {A.~M.}\ \bibnamefont
  {Popov}}, \bibinfo {author} {\bibfnamefont {I.~V.}\ \bibnamefont {Lebedeva}},
  \bibinfo {author} {\bibfnamefont {A.~A.}\ \bibnamefont {Knizhnik}}, \bibinfo
  {author} {\bibfnamefont {Yu.~E.}\ \bibnamefont {Lozovik}}, \ and\ \bibinfo
  {author} {\bibfnamefont {B.~V.}\ \bibnamefont {Potapkin}},\ }\bibfield
  {title} {\enquote {\bibinfo {title} {Commensurate-incommensurate phase
  transition in bilayer graphene},}\ }\href {\doibase
  10.1103/PhysRevB.84.045404} {\bibfield  {journal} {\bibinfo  {journal} {Phys.
  Rev. B}\ }\textbf {\bibinfo {volume} {84}},\ \bibinfo {pages} {045404}
  (\bibinfo {year} {2011})}\BibitemShut {NoStop}%
\bibitem [{\citenamefont {Lebedev}\ \emph {et~al.}(2016)\citenamefont
  {Lebedev}, \citenamefont {Lebedeva}, \citenamefont {Knizhnik},\ and\
  \citenamefont {Popov}}]{Lebedev2015}%
  \BibitemOpen
  \bibfield  {author} {\bibinfo {author} {\bibfnamefont {A.~V.}\ \bibnamefont
  {Lebedev}}, \bibinfo {author} {\bibfnamefont {I.~V.}\ \bibnamefont
  {Lebedeva}}, \bibinfo {author} {\bibfnamefont {A.~A.}\ \bibnamefont
  {Knizhnik}}, \ and\ \bibinfo {author} {\bibfnamefont {A.~M.}\ \bibnamefont
  {Popov}},\ }\bibfield  {title} {\enquote {\bibinfo {title} {Interlayer
  interaction and related properties of bilayer hexagonal boron nitride:
  \textit{ab initio} study},}\ }\href {\doibase 10.1039/C5RA20882C} {\bibfield
  {journal} {\bibinfo  {journal} {RSC Advances}\ }\textbf {\bibinfo {volume}
  {6}},\ \bibinfo {pages} {6423--6435} (\bibinfo {year} {2016})}\BibitemShut
  {NoStop}%
\bibitem [{\citenamefont {Hattendorf}\ \emph {et~al.}(2013)\citenamefont
  {Hattendorf}, \citenamefont {Georgi}, \citenamefont {Liebmann},\ and\
  \citenamefont {Morgenstern}}]{Hattendorf2013}%
  \BibitemOpen
  \bibfield  {author} {\bibinfo {author} {\bibfnamefont {S.}~\bibnamefont
  {Hattendorf}}, \bibinfo {author} {\bibfnamefont {A.}~\bibnamefont {Georgi}},
  \bibinfo {author} {\bibfnamefont {M.}~\bibnamefont {Liebmann}}, \ and\
  \bibinfo {author} {\bibfnamefont {M.}~\bibnamefont {Morgenstern}},\
  }\bibfield  {title} {\enquote {\bibinfo {title} {Networks of {ABA} and {ABC}
  stacked graphene on mica observed by scanning tunneling microscopy},}\ }\href
  {\doibase 10.1016/j.susc.2013.01.005} {\bibfield  {journal} {\bibinfo
  {journal} {Surf. Sci.}\ }\textbf {\bibinfo {volume} {610}},\ \bibinfo {pages}
  {53--58} (\bibinfo {year} {2013})}\BibitemShut {NoStop}%
\bibitem [{\citenamefont {San-Jose}\ \emph {et~al.}(2014)\citenamefont
  {San-Jose}, \citenamefont {Gorbachev}, \citenamefont {Geim}, \citenamefont
  {Novoselov},\ and\ \citenamefont {Guinea}}]{San-Jose2014}%
  \BibitemOpen
  \bibfield  {author} {\bibinfo {author} {\bibfnamefont {P.}~\bibnamefont
  {San-Jose}}, \bibinfo {author} {\bibfnamefont {R.~V.}\ \bibnamefont
  {Gorbachev}}, \bibinfo {author} {\bibfnamefont {A.~K.}\ \bibnamefont {Geim}},
  \bibinfo {author} {\bibfnamefont {K.~S.}\ \bibnamefont {Novoselov}}, \ and\
  \bibinfo {author} {\bibfnamefont {F.}~\bibnamefont {Guinea}},\ }\bibfield
  {title} {\enquote {\bibinfo {title} {Stacking boundaries and transport in
  bilayer graphene},}\ }\href {\doibase 10.1021/nl500230a} {\bibfield
  {journal} {\bibinfo  {journal} {Nano Lett.}\ }\textbf {\bibinfo {volume}
  {14}},\ \bibinfo {pages} {2052--2057} (\bibinfo {year} {2014})}\BibitemShut
  {NoStop}%
\bibitem [{\citenamefont {Lalmi}\ \emph {et~al.}(2014)\citenamefont {Lalmi},
  \citenamefont {Girard}, \citenamefont {Pallecchi}, \citenamefont {Silly},
  \citenamefont {David}, \citenamefont {Latil}, \citenamefont {Sirotti},\ and\
  \citenamefont {Ouerghi}}]{Lalmi2014}%
  \BibitemOpen
  \bibfield  {author} {\bibinfo {author} {\bibfnamefont {B.}~\bibnamefont
  {Lalmi}}, \bibinfo {author} {\bibfnamefont {J.~C.}\ \bibnamefont {Girard}},
  \bibinfo {author} {\bibfnamefont {E.}~\bibnamefont {Pallecchi}}, \bibinfo
  {author} {\bibfnamefont {M.}~\bibnamefont {Silly}}, \bibinfo {author}
  {\bibfnamefont {C.}~\bibnamefont {David}}, \bibinfo {author} {\bibfnamefont
  {S.}~\bibnamefont {Latil}}, \bibinfo {author} {\bibfnamefont
  {F.}~\bibnamefont {Sirotti}}, \ and\ \bibinfo {author} {\bibfnamefont
  {A.}~\bibnamefont {Ouerghi}},\ }\bibfield  {title} {\enquote {\bibinfo
  {title} {Flower-shaped domains and wrinkles in trilayer epitaxial graphene on
  silicon carbide},}\ }\href {\doibase 10.1038/srep04066} {\bibfield  {journal}
  {\bibinfo  {journal} {Sci. Rep.}\ }\textbf {\bibinfo {volume} {4}},\ \bibinfo
  {pages} {4066} (\bibinfo {year} {2014})}\BibitemShut {NoStop}%
\bibitem [{\citenamefont {Benameur}\ \emph {et~al.}(2015)\citenamefont
  {Benameur}, \citenamefont {Gargiulo}, \citenamefont {Manzeli}, \citenamefont
  {Aut\`{e}s}, \citenamefont {Tosun}, \citenamefont {Yazyev},\ and\
  \citenamefont {Kis}}]{Benameur2015}%
  \BibitemOpen
  \bibfield  {author} {\bibinfo {author} {\bibfnamefont {M.~M.}\ \bibnamefont
  {Benameur}}, \bibinfo {author} {\bibfnamefont {F.}~\bibnamefont {Gargiulo}},
  \bibinfo {author} {\bibfnamefont {S.}~\bibnamefont {Manzeli}}, \bibinfo
  {author} {\bibfnamefont {G.}~\bibnamefont {Aut\`{e}s}}, \bibinfo {author}
  {\bibfnamefont {M.}~\bibnamefont {Tosun}}, \bibinfo {author} {\bibfnamefont
  {O.~V.}\ \bibnamefont {Yazyev}}, \ and\ \bibinfo {author} {\bibfnamefont
  {A.}~\bibnamefont {Kis}},\ }\bibfield  {title} {\enquote {\bibinfo {title}
  {Electromechanical oscillations in bilayer graphene},}\ }\href {\doibase
  10.1038/ncomms9582} {\bibfield  {journal} {\bibinfo  {journal} {Nat. Comm.}\
  }\textbf {\bibinfo {volume} {6}},\ \bibinfo {pages} {8582} (\bibinfo {year}
  {2015})}\BibitemShut {NoStop}%
\bibitem [{\citenamefont {Koshino}(2013)}]{Koshino2013}%
  \BibitemOpen
  \bibfield  {author} {\bibinfo {author} {\bibfnamefont {M.}~\bibnamefont
  {Koshino}},\ }\bibfield  {title} {\enquote {\bibinfo {title} {Electronic
  transmission through {AB-BA} domain boundary in bilayer graphene},}\ }\href
  {\doibase 10.1103/PhysRevB.88.115409} {\bibfield  {journal} {\bibinfo
  {journal} {Phys. Rev. B}\ }\textbf {\bibinfo {volume} {88}},\ \bibinfo
  {pages} {115409} (\bibinfo {year} {2013})}\BibitemShut {NoStop}%
\bibitem [{\citenamefont {Gong}\ \emph {et~al.}(2013)\citenamefont {Gong},
  \citenamefont {Young}, \citenamefont {Kinloch}, \citenamefont {Haigh},
  \citenamefont {Warner}, \citenamefont {Hinks}, \citenamefont {Xu},
  \citenamefont {Li}, \citenamefont {Ding}, \citenamefont {Riaz}, \citenamefont
  {Jalil},\ and\ \citenamefont {Novoselov}}]{Gong2013}%
  \BibitemOpen
  \bibfield  {author} {\bibinfo {author} {\bibfnamefont {L.}~\bibnamefont
  {Gong}}, \bibinfo {author} {\bibfnamefont {R.~J.}\ \bibnamefont {Young}},
  \bibinfo {author} {\bibfnamefont {I.~A.}\ \bibnamefont {Kinloch}}, \bibinfo
  {author} {\bibfnamefont {S.~J.}\ \bibnamefont {Haigh}}, \bibinfo {author}
  {\bibfnamefont {J.~H.}\ \bibnamefont {Warner}}, \bibinfo {author}
  {\bibfnamefont {J.~A.}\ \bibnamefont {Hinks}}, \bibinfo {author}
  {\bibfnamefont {Z.}~\bibnamefont {Xu}}, \bibinfo {author} {\bibfnamefont
  {L.}~\bibnamefont {Li}}, \bibinfo {author} {\bibfnamefont {F.}~\bibnamefont
  {Ding}}, \bibinfo {author} {\bibfnamefont {I.}~\bibnamefont {Riaz}}, \bibinfo
  {author} {\bibfnamefont {R.}~\bibnamefont {Jalil}}, \ and\ \bibinfo {author}
  {\bibfnamefont {K.~S.}\ \bibnamefont {Novoselov}},\ }\bibfield  {title}
  {\enquote {\bibinfo {title} {Reversible loss of {Bernal} stacking during the
  deformation of few-layer graphene in nanocomposites},}\ }\href {\doibase
  10.1021/nn402830f} {\bibfield  {journal} {\bibinfo  {journal} {ACS Nano}\
  }\textbf {\bibinfo {volume} {7}},\ \bibinfo {pages} {7287--7294} (\bibinfo
  {year} {2013})}\BibitemShut {NoStop}%
\bibitem [{\citenamefont {Bichoutskaia}\ \emph {et~al.}(2006)\citenamefont
  {Bichoutskaia}, \citenamefont {Heggie}, \citenamefont {Lozovik},\ and\
  \citenamefont {Popov}}]{Bichoutskaia2006}%
  \BibitemOpen
  \bibfield  {author} {\bibinfo {author} {\bibfnamefont {E.}~\bibnamefont
  {Bichoutskaia}}, \bibinfo {author} {\bibfnamefont {M.~I.}\ \bibnamefont
  {Heggie}}, \bibinfo {author} {\bibfnamefont {Yu.~E.}\ \bibnamefont
  {Lozovik}}, \ and\ \bibinfo {author} {\bibfnamefont {A.~M.}\ \bibnamefont
  {Popov}},\ }\bibfield  {title} {\enquote {\bibinfo {title} {Multi-walled
  nanotubes: Commensurate-incommensurate phase transition and {NEMS}
  applications},}\ }\href {\doibase 10.1080/15363830600663412} {\bibfield
  {journal} {\bibinfo  {journal} {Fullerenes, Nanotubes, Carbon Nanostruct.}\
  }\textbf {\bibinfo {volume} {14}},\ \bibinfo {pages} {131--140} (\bibinfo
  {year} {2006})}\BibitemShut {NoStop}%
\bibitem [{\citenamefont {Warner}\ \emph {et~al.}(2010)\citenamefont {Warner},
  \citenamefont {R\"{u}mmeli}, \citenamefont {Bachmatiuk},\ and\ \citenamefont
  {B\"{u}chner}}]{Warner2010}%
  \BibitemOpen
  \bibfield  {author} {\bibinfo {author} {\bibfnamefont {J.~H.}\ \bibnamefont
  {Warner}}, \bibinfo {author} {\bibfnamefont {M.~H.}\ \bibnamefont
  {R\"{u}mmeli}}, \bibinfo {author} {\bibfnamefont {A.}~\bibnamefont
  {Bachmatiuk}}, \ and\ \bibinfo {author} {\bibfnamefont {B.}~\bibnamefont
  {B\"{u}chner}},\ }\bibfield  {title} {\enquote {\bibinfo {title} {Atomic
  resolution imaging and topography of boron nitride sheets produced by
  chemical exfoliation},}\ }\href {\doibase 10.1021/nn901648q} {\bibfield
  {journal} {\bibinfo  {journal} {ACS Nano}\ }\textbf {\bibinfo {volume} {4}},\
  \bibinfo {pages} {1299--1304} (\bibinfo {year} {2010})}\BibitemShut {NoStop}%
\bibitem [{\citenamefont {Porovski\u{i}}\ and\ \citenamefont
  {Talapov}(1978)}]{Pokrovsky1978}%
  \BibitemOpen
  \bibfield  {author} {\bibinfo {author} {\bibfnamefont {V.~L.}\ \bibnamefont
  {Porovski\u{i}}}\ and\ \bibinfo {author} {\bibfnamefont {A.~L.}\ \bibnamefont
  {Talapov}},\ }\bibfield  {title} {\enquote {\bibinfo {title} {Phase
  transitions and vibrational spectra of almost commensurate structures},}\
  }\href@noop {} {\bibfield  {journal} {\bibinfo  {journal} {Soviet Physics
  JETP}\ }\textbf {\bibinfo {volume} {48}},\ \bibinfo {pages} {579--582}
  (\bibinfo {year} {1978})}\BibitemShut {NoStop}%
\bibitem [{\citenamefont {Popov}\ \emph {et~al.}(2009)\citenamefont {Popov},
  \citenamefont {Lozovik}, \citenamefont {Sobennikov},\ and\ \citenamefont
  {Knizhnik}}]{Popov2009}%
  \BibitemOpen
  \bibfield  {author} {\bibinfo {author} {\bibfnamefont {A.~M.}\ \bibnamefont
  {Popov}}, \bibinfo {author} {\bibfnamefont {Y.~E.}\ \bibnamefont {Lozovik}},
  \bibinfo {author} {\bibfnamefont {A.~S.}\ \bibnamefont {Sobennikov}}, \ and\
  \bibinfo {author} {\bibfnamefont {A.~A.}\ \bibnamefont {Knizhnik}},\
  }\bibfield  {title} {\enquote {\bibinfo {title} {Nanomechanical properties
  and phase transitions in double-walled carbon nanotube (5,5)@(10,10):
  \textit{Ab initio} calculations},}\ }\href {\doibase
  10.1134/S1063776109040104} {\bibfield  {journal} {\bibinfo  {journal} {JETP}\
  }\textbf {\bibinfo {volume} {108}},\ \bibinfo {pages} {621--628} (\bibinfo
  {year} {2009})}\BibitemShut {NoStop}%
\bibitem [{\citenamefont {Woods}\ \emph {et~al.}(2014)\citenamefont {Woods},
  \citenamefont {Britnell}, \citenamefont {Eckmann}, \citenamefont {Ma},
  \citenamefont {Lu}, \citenamefont {Guo}, \citenamefont {Lin}, \citenamefont
  {Yu}, \citenamefont {Cao}, \citenamefont {Gorbachev}, \citenamefont
  {Kretinin}, \citenamefont {Park}, \citenamefont {Ponomarenko}, \citenamefont
  {Katsnelson}, \citenamefont {Gornostyrev}, \citenamefont {Watanabe},
  \citenamefont {Taniguchi}, \citenamefont {Casiraghi}, \citenamefont {Gao},
  \citenamefont {Geim},\ and\ \citenamefont {Novoselov}}]{Woods2014}%
  \BibitemOpen
  \bibfield  {author} {\bibinfo {author} {\bibfnamefont {C.~R.}\ \bibnamefont
  {Woods}}, \bibinfo {author} {\bibfnamefont {L.}~\bibnamefont {Britnell}},
  \bibinfo {author} {\bibfnamefont {A.}~\bibnamefont {Eckmann}}, \bibinfo
  {author} {\bibfnamefont {R.~S.}\ \bibnamefont {Ma}}, \bibinfo {author}
  {\bibfnamefont {J.~C.}\ \bibnamefont {Lu}}, \bibinfo {author} {\bibfnamefont
  {H.~M.}\ \bibnamefont {Guo}}, \bibinfo {author} {\bibfnamefont
  {X.}~\bibnamefont {Lin}}, \bibinfo {author} {\bibfnamefont {G.~L.}\
  \bibnamefont {Yu}}, \bibinfo {author} {\bibfnamefont {Y.}~\bibnamefont
  {Cao}}, \bibinfo {author} {\bibfnamefont {R.~V.}\ \bibnamefont {Gorbachev}},
  \bibinfo {author} {\bibfnamefont {A.~V.}\ \bibnamefont {Kretinin}}, \bibinfo
  {author} {\bibfnamefont {J.}~\bibnamefont {Park}}, \bibinfo {author}
  {\bibfnamefont {L.~A.}\ \bibnamefont {Ponomarenko}}, \bibinfo {author}
  {\bibfnamefont {M.~I.}\ \bibnamefont {Katsnelson}}, \bibinfo {author}
  {\bibfnamefont {Yu.~N.}\ \bibnamefont {Gornostyrev}}, \bibinfo {author}
  {\bibfnamefont {K.}~\bibnamefont {Watanabe}}, \bibinfo {author}
  {\bibfnamefont {T.}~\bibnamefont {Taniguchi}}, \bibinfo {author}
  {\bibfnamefont {C.}~\bibnamefont {Casiraghi}}, \bibinfo {author}
  {\bibfnamefont {H.-J.}\ \bibnamefont {Gao}}, \bibinfo {author} {\bibfnamefont
  {A.~K.}\ \bibnamefont {Geim}}, \ and\ \bibinfo {author} {\bibfnamefont
  {K.~S.}\ \bibnamefont {Novoselov}},\ }\bibfield  {title} {\enquote {\bibinfo
  {title} {Commensurate-incommensurate transition in graphene on hexagonal
  boron nitride},}\ }\href {\doibase 10.1038/nphys2954} {\bibfield  {journal}
  {\bibinfo  {journal} {Nature Physics}\ }\textbf {\bibinfo {volume} {10}},\
  \bibinfo {pages} {451--456} (\bibinfo {year} {2014})}\BibitemShut {NoStop}%
\bibitem [{\citenamefont {Korhonen}\ and\ \citenamefont
  {Koskinen}(2015)}]{Korhonen2015}%
  \BibitemOpen
  \bibfield  {author} {\bibinfo {author} {\bibfnamefont {T.}~\bibnamefont
  {Korhonen}}\ and\ \bibinfo {author} {\bibfnamefont {P.}~\bibnamefont
  {Koskinen}},\ }\bibfield  {title} {\enquote {\bibinfo {title} {Peeling of
  multilayer graphene creates complex interlayer sliding patterns},}\ }\href
  {\doibase 10.1103/PhysRevB.92.115427} {\bibfield  {journal} {\bibinfo
  {journal} {Phys. Rev. B}\ }\textbf {\bibinfo {volume} {92}},\ \bibinfo
  {pages} {115427} (\bibinfo {year} {2015})}\BibitemShut {NoStop}%
\bibitem [{\citenamefont {Lee}\ \emph {et~al.}(2010)\citenamefont {Lee},
  \citenamefont {Murray}, \citenamefont {Kong}, \citenamefont {Lundqvist},\
  and\ \citenamefont {Langreth}}]{Lee2010}%
  \BibitemOpen
  \bibfield  {author} {\bibinfo {author} {\bibfnamefont {K.}~\bibnamefont
  {Lee}}, \bibinfo {author} {\bibfnamefont {E.~D.}\ \bibnamefont {Murray}},
  \bibinfo {author} {\bibfnamefont {L.}~\bibnamefont {Kong}}, \bibinfo {author}
  {\bibfnamefont {B.~I.}\ \bibnamefont {Lundqvist}}, \ and\ \bibinfo {author}
  {\bibfnamefont {D.~C.}\ \bibnamefont {Langreth}},\ }\bibfield  {title}
  {\enquote {\bibinfo {title} {Higher-accuracy van der waals density
  functional},}\ }\href {\doibase 10.1103/PhysRevB.82.081101} {\bibfield
  {journal} {\bibinfo  {journal} {Phys. Rev. B}\ }\textbf {\bibinfo {volume}
  {82}},\ \bibinfo {pages} {081101} (\bibinfo {year} {2010})}\BibitemShut
  {NoStop}%
\bibitem [{\citenamefont {Kresse}\ and\ \citenamefont
  {Furthm\"{u}ller}(1996)}]{Kresse1996}%
  \BibitemOpen
  \bibfield  {author} {\bibinfo {author} {\bibfnamefont {G.}~\bibnamefont
  {Kresse}}\ and\ \bibinfo {author} {\bibfnamefont {J.}~\bibnamefont
  {Furthm\"{u}ller}},\ }\bibfield  {title} {\enquote {\bibinfo {title}
  {Efficient iterative schemes for \textit{ab initio} total-energy calculations
  using a plane-wave basis set},}\ }\href {\doibase 10.1103/PhysRevB.54.11169}
  {\bibfield  {journal} {\bibinfo  {journal} {Phys. Rev. B}\ }\textbf {\bibinfo
  {volume} {54}},\ \bibinfo {pages} {11169--11186} (\bibinfo {year}
  {1996})}\BibitemShut {NoStop}%
\bibitem [{\citenamefont {Kresse}\ and\ \citenamefont
  {Joubert}(1999)}]{Kresse1999}%
  \BibitemOpen
  \bibfield  {author} {\bibinfo {author} {\bibfnamefont {G.}~\bibnamefont
  {Kresse}}\ and\ \bibinfo {author} {\bibfnamefont {D.}~\bibnamefont
  {Joubert}},\ }\bibfield  {title} {\enquote {\bibinfo {title} {From ultrasoft
  pseudopotentials to the projector augmented-wave method},}\ }\href {\doibase
  10.1103/PhysRevB.59.1758} {\bibfield  {journal} {\bibinfo  {journal} {Phys.
  Rev. B}\ }\textbf {\bibinfo {volume} {59}},\ \bibinfo {pages} {1758--1775}
  (\bibinfo {year} {1999})}\BibitemShut {NoStop}%
\bibitem [{\citenamefont {Monkhorst}\ and\ \citenamefont
  {Pack}(1976)}]{Monkhorst1976}%
  \BibitemOpen
  \bibfield  {author} {\bibinfo {author} {\bibfnamefont {H.~J.}\ \bibnamefont
  {Monkhorst}}\ and\ \bibinfo {author} {\bibfnamefont {J.~D.}\ \bibnamefont
  {Pack}},\ }\bibfield  {title} {\enquote {\bibinfo {title} {Special points for
  {Brillouin}-zone integrations},}\ }\href {\doibase 10.1103/PhysRevB.13.5188}
  {\bibfield  {journal} {\bibinfo  {journal} {Phys. Rev. B}\ }\textbf {\bibinfo
  {volume} {13}},\ \bibinfo {pages} {5188--5192} (\bibinfo {year}
  {1976})}\BibitemShut {NoStop}%
\bibitem [{\citenamefont {Lebedeva}\ \emph
  {et~al.}(2011{\natexlab{a}})\citenamefont {Lebedeva}, \citenamefont
  {Knizhnik}, \citenamefont {Popov}, \citenamefont {Lozovik},\ and\
  \citenamefont {Potapkin}}]{Lebedeva2011}%
  \BibitemOpen
  \bibfield  {author} {\bibinfo {author} {\bibfnamefont {I.~V.}\ \bibnamefont
  {Lebedeva}}, \bibinfo {author} {\bibfnamefont {A.~A.}\ \bibnamefont
  {Knizhnik}}, \bibinfo {author} {\bibfnamefont {A.~M.}\ \bibnamefont {Popov}},
  \bibinfo {author} {\bibfnamefont {Yu.~E.}\ \bibnamefont {Lozovik}}, \ and\
  \bibinfo {author} {\bibfnamefont {B.~V.}\ \bibnamefont {Potapkin}},\
  }\bibfield  {title} {\enquote {\bibinfo {title} {Interlayer interaction and
  relative vibrations of bilayer graphene},}\ }\href {\doibase
  10.1039/c0cp02614j} {\bibfield  {journal} {\bibinfo  {journal} {Phys. Chem.
  Chem. Phys.}\ }\textbf {\bibinfo {volume} {13}},\ \bibinfo {pages}
  {5687--5695} (\bibinfo {year} {2011}{\natexlab{a}})}\BibitemShut {NoStop}%
\bibitem [{\citenamefont {Bernal}(1924)}]{Bernal1924}%
  \BibitemOpen
  \bibfield  {author} {\bibinfo {author} {\bibfnamefont {J.~D.}\ \bibnamefont
  {Bernal}},\ }\bibfield  {title} {\enquote {\bibinfo {title} {The structure of
  graphite},}\ }\href@noop {} {\bibfield  {journal} {\bibinfo  {journal} {Proc.
  R. Soc. London, Ser. A}\ }\textbf {\bibinfo {volume} {106}},\ \bibinfo
  {pages} {749--773} (\bibinfo {year} {1924})}\BibitemShut {NoStop}%
\bibitem [{\citenamefont {Baskin}\ and\ \citenamefont
  {Meyer}(1955)}]{Baskin1955}%
  \BibitemOpen
  \bibfield  {author} {\bibinfo {author} {\bibfnamefont {V.}~\bibnamefont
  {Baskin}}\ and\ \bibinfo {author} {\bibfnamefont {L.}~\bibnamefont {Meyer}},\
  }\bibfield  {title} {\enquote {\bibinfo {title} {Lattice constants of
  graphite at low temperatures},}\ }\href {\doibase 10.1103/PhysRev.100.544}
  {\bibfield  {journal} {\bibinfo  {journal} {Phys. Rev.}\ }\textbf {\bibinfo
  {volume} {100}},\ \bibinfo {pages} {544} (\bibinfo {year}
  {1955})}\BibitemShut {NoStop}%
\bibitem [{\citenamefont {Wyckoff}(1963)}]{Wyckoff1963}%
  \BibitemOpen
  \bibfield  {author} {\bibinfo {author} {\bibfnamefont {R.~W.~G.}\
  \bibnamefont {Wyckoff}},\ }\href@noop {} {\emph {\bibinfo {title} {Crystal
  Structures (2nd Edition)}}},\ Vol.~\bibinfo {volume} {1}\ (\bibinfo
  {publisher} {HarperCollins, New York, USA},\ \bibinfo {year}
  {1963})\BibitemShut {NoStop}%
\bibitem [{\citenamefont {Lynch}\ and\ \citenamefont
  {Drickamer}(1966)}]{Lynch1966}%
  \BibitemOpen
  \bibfield  {author} {\bibinfo {author} {\bibfnamefont {R.~W.}\ \bibnamefont
  {Lynch}}\ and\ \bibinfo {author} {\bibfnamefont {H.~G.}\ \bibnamefont
  {Drickamer}},\ }\bibfield  {title} {\enquote {\bibinfo {title} {Effect of
  high pressure on the lattice parameters of diamond, graphite, and hexagonal
  boron nitride},}\ }\href {\doibase 10.1063/1.1726442} {\bibfield  {journal}
  {\bibinfo  {journal} {J. Chem. Phys.}\ }\textbf {\bibinfo {volume} {44}},\
  \bibinfo {pages} {181--184} (\bibinfo {year} {1966})}\BibitemShut {NoStop}%
\bibitem [{\citenamefont {Ludsteck}(1972)}]{Ludsteck1972}%
  \BibitemOpen
  \bibfield  {author} {\bibinfo {author} {\bibfnamefont {Von~A.}\ \bibnamefont
  {Ludsteck}},\ }\bibfield  {title} {\enquote {\bibinfo {title} {Bestimmung der
  {\aa}nderung der gitterkonstanten und des anisotropen debye–waller-faktors
  von graphit mittels neutronenbeugung im temperaturbereich von 25 bis
  1850$^{\circ}$ {C}},}\ }\href {\doibase 10.1107/S0567739472000130} {\bibfield
   {journal} {\bibinfo  {journal} {Acta Crystallographica, Section A}\ }\textbf
  {\bibinfo {volume} {28}},\ \bibinfo {pages} {59--65} (\bibinfo {year}
  {1972})}\BibitemShut {NoStop}%
\bibitem [{\citenamefont {Trucano}\ and\ \citenamefont
  {Chen}(1975)}]{Trucano1975}%
  \BibitemOpen
  \bibfield  {author} {\bibinfo {author} {\bibfnamefont {P.}~\bibnamefont
  {Trucano}}\ and\ \bibinfo {author} {\bibfnamefont {R.}~\bibnamefont {Chen}},\
  }\bibfield  {title} {\enquote {\bibinfo {title} {Structure of graphite by
  neutron diffraction},}\ }\href {\doibase 10.1038/258136a0} {\bibfield
  {journal} {\bibinfo  {journal} {Nature}\ }\textbf {\bibinfo {volume} {258}},\
  \bibinfo {pages} {136--137} (\bibinfo {year} {1975})}\BibitemShut {NoStop}%
\bibitem [{\citenamefont {Zhao}\ and\ \citenamefont {Spain}(1989)}]{Zhao1989}%
  \BibitemOpen
  \bibfield  {author} {\bibinfo {author} {\bibfnamefont {Y.~X.}\ \bibnamefont
  {Zhao}}\ and\ \bibinfo {author} {\bibfnamefont {I.~L.}\ \bibnamefont
  {Spain}},\ }\bibfield  {title} {\enquote {\bibinfo {title} {X-ray diffraction
  data for graphite to 20 {GPa}},}\ }\href {\doibase 10.1103/PhysRevB.40.993}
  {\bibfield  {journal} {\bibinfo  {journal} {Phys. Rev. B}\ }\textbf {\bibinfo
  {volume} {40}},\ \bibinfo {pages} {993--997} (\bibinfo {year}
  {1989})}\BibitemShut {NoStop}%
\bibitem [{\citenamefont {Bosak}\ \emph {et~al.}(2007)\citenamefont {Bosak},
  \citenamefont {Krisch}, \citenamefont {Mohr}, \citenamefont {Maultzsch},\
  and\ \citenamefont {Thomsen}}]{Bosak2007}%
  \BibitemOpen
  \bibfield  {author} {\bibinfo {author} {\bibfnamefont {A.}~\bibnamefont
  {Bosak}}, \bibinfo {author} {\bibfnamefont {M.}~\bibnamefont {Krisch}},
  \bibinfo {author} {\bibfnamefont {M.}~\bibnamefont {Mohr}}, \bibinfo {author}
  {\bibfnamefont {J.}~\bibnamefont {Maultzsch}}, \ and\ \bibinfo {author}
  {\bibfnamefont {C.}~\bibnamefont {Thomsen}},\ }\bibfield  {title} {\enquote
  {\bibinfo {title} {Elasticity of single-crystalline graphite: Inelastic x-ray
  scattering study},}\ }\href {\doibase 10.1103/PhysRevB.75.153408} {\bibfield
  {journal} {\bibinfo  {journal} {Phys. Rev. B}\ }\textbf {\bibinfo {volume}
  {75}},\ \bibinfo {pages} {153408} (\bibinfo {year} {2007})}\BibitemShut
  {NoStop}%
\bibitem [{\citenamefont {Pease}(1950)}]{Pease1950}%
  \BibitemOpen
  \bibfield  {author} {\bibinfo {author} {\bibfnamefont {R.~S.}\ \bibnamefont
  {Pease}},\ }\bibfield  {title} {\enquote {\bibinfo {title} {Crystal structure
  of boron nitride},}\ }\href {\doibase 10.1038/165722b0} {\bibfield  {journal}
  {\bibinfo  {journal} {Nature}\ }\textbf {\bibinfo {volume} {165}},\ \bibinfo
  {pages} {722--723} (\bibinfo {year} {1950})}\BibitemShut {NoStop}%
\bibitem [{\citenamefont {Pease}(1952)}]{Pease1952}%
  \BibitemOpen
  \bibfield  {author} {\bibinfo {author} {\bibfnamefont {R.~S.}\ \bibnamefont
  {Pease}},\ }\bibfield  {title} {\enquote {\bibinfo {title} {An x-ray study of
  boron nitride},}\ }\href {\doibase 10.1107/S0365110X52001064} {\bibfield
  {journal} {\bibinfo  {journal} {Acta Crystallographica}\ }\textbf {\bibinfo
  {volume} {5}},\ \bibinfo {pages} {356--361} (\bibinfo {year}
  {1952})}\BibitemShut {NoStop}%
\bibitem [{\citenamefont {Solozhenko}\ \emph {et~al.}(1995)\citenamefont
  {Solozhenko}, \citenamefont {Will},\ and\ \citenamefont
  {Elf}}]{Solozhenko1995}%
  \BibitemOpen
  \bibfield  {author} {\bibinfo {author} {\bibfnamefont {V.~L.}\ \bibnamefont
  {Solozhenko}}, \bibinfo {author} {\bibfnamefont {G.}~\bibnamefont {Will}}, \
  and\ \bibinfo {author} {\bibfnamefont {F.}~\bibnamefont {Elf}},\ }\bibfield
  {title} {\enquote {\bibinfo {title} {Isothermal compression of hexagonal
  graphite-like boron nitride up to 12 {GPa}},}\ }\href {\doibase
  10.1016/0038-1098(95)00381-9} {\bibfield  {journal} {\bibinfo  {journal}
  {Solid State Communications}\ }\textbf {\bibinfo {volume} {96}},\ \bibinfo
  {pages} {1--3} (\bibinfo {year} {1995})}\BibitemShut {NoStop}%
\bibitem [{\citenamefont {Solozhenko}\ and\ \citenamefont
  {Peun}(1997)}]{Solozhenko1997}%
  \BibitemOpen
  \bibfield  {author} {\bibinfo {author} {\bibfnamefont {V.~L.}\ \bibnamefont
  {Solozhenko}}\ and\ \bibinfo {author} {\bibfnamefont {T.}~\bibnamefont
  {Peun}},\ }\bibfield  {title} {\enquote {\bibinfo {title} {Compression and
  thermal expansion of hexagonal graphite-like boron nitride up to 7 {GPa} and
  1800 {K}},}\ }\href {\doibase 10.1016/S0022-3697(97)00037-1} {\bibfield
  {journal} {\bibinfo  {journal} {J. Phys. Chem. Solids}\ }\textbf {\bibinfo
  {volume} {58}},\ \bibinfo {pages} {1321--1323} (\bibinfo {year}
  {1997})}\BibitemShut {NoStop}%
\bibitem [{\citenamefont {Solozhenko}\ \emph {et~al.}(2001)\citenamefont
  {Solozhenko}, \citenamefont {Lazarenko}, \citenamefont {Petitet},\ and\
  \citenamefont {Kanaev}}]{Solozhenko2001}%
  \BibitemOpen
  \bibfield  {author} {\bibinfo {author} {\bibfnamefont {V.~L.}\ \bibnamefont
  {Solozhenko}}, \bibinfo {author} {\bibfnamefont {A.~G.}\ \bibnamefont
  {Lazarenko}}, \bibinfo {author} {\bibfnamefont {J.-P.}\ \bibnamefont
  {Petitet}}, \ and\ \bibinfo {author} {\bibfnamefont {A.~V.}\ \bibnamefont
  {Kanaev}},\ }\bibfield  {title} {\enquote {\bibinfo {title} {Bandgap energy
  of graphite-like hexagonal boron nitride},}\ }\href {\doibase
  10.1016/S0022-3697(01)00030-0} {\bibfield  {journal} {\bibinfo  {journal} {J.
  Phys. Chem. Solids}\ }\textbf {\bibinfo {volume} {62}},\ \bibinfo {pages}
  {1331--1334} (\bibinfo {year} {2001})}\BibitemShut {NoStop}%
\bibitem [{\citenamefont {Paszkowicz}\ \emph {et~al.}(2002)\citenamefont
  {Paszkowicz}, \citenamefont {Pelka}, \citenamefont {Knapp}, \citenamefont
  {Szyszko},\ and\ \citenamefont {Podsiadlo}}]{Paszkowicz2002}%
  \BibitemOpen
  \bibfield  {author} {\bibinfo {author} {\bibfnamefont {W.}~\bibnamefont
  {Paszkowicz}}, \bibinfo {author} {\bibfnamefont {J.~B.}\ \bibnamefont
  {Pelka}}, \bibinfo {author} {\bibfnamefont {M.}~\bibnamefont {Knapp}},
  \bibinfo {author} {\bibfnamefont {T.}~\bibnamefont {Szyszko}}, \ and\
  \bibinfo {author} {\bibfnamefont {S.}~\bibnamefont {Podsiadlo}},\ }\bibfield
  {title} {\enquote {\bibinfo {title} {Lattice parameters and anisotropic
  thermal expansion of hexagonal boron nitride in the 10--297.5 {K} temperature
  range},}\ }\href {\doibase 10.1007/s003390100999} {\bibfield  {journal}
  {\bibinfo  {journal} {Appl. Phys. A}\ }\textbf {\bibinfo {volume} {75}},\
  \bibinfo {pages} {431--435} (\bibinfo {year} {2002})}\BibitemShut {NoStop}%
\bibitem [{\citenamefont {Bosak}\ \emph {et~al.}(2006)\citenamefont {Bosak},
  \citenamefont {Serrano}, \citenamefont {Krisch}, \citenamefont {Watanabe},
  \citenamefont {Taniguchi},\ and\ \citenamefont {Kanda}}]{Bosak2006}%
  \BibitemOpen
  \bibfield  {author} {\bibinfo {author} {\bibfnamefont {A.}~\bibnamefont
  {Bosak}}, \bibinfo {author} {\bibfnamefont {J.}~\bibnamefont {Serrano}},
  \bibinfo {author} {\bibfnamefont {M.}~\bibnamefont {Krisch}}, \bibinfo
  {author} {\bibfnamefont {K.}~\bibnamefont {Watanabe}}, \bibinfo {author}
  {\bibfnamefont {T.}~\bibnamefont {Taniguchi}}, \ and\ \bibinfo {author}
  {\bibfnamefont {H.}~\bibnamefont {Kanda}},\ }\bibfield  {title} {\enquote
  {\bibinfo {title} {Elasticity of hexagonal boron nitride: Inelastic x-ray
  scattering measurements},}\ }\href {\doibase 10.1103/PhysRevB.73.041402}
  {\bibfield  {journal} {\bibinfo  {journal} {Phys. Rev. B}\ }\textbf {\bibinfo
  {volume} {73}},\ \bibinfo {pages} {041402} (\bibinfo {year}
  {2006})}\BibitemShut {NoStop}%
\bibitem [{\citenamefont {Fuchizaki}\ \emph {et~al.}(2008)\citenamefont
  {Fuchizaki}, \citenamefont {Nakamichi}, \citenamefont {Saitoh},\ and\
  \citenamefont {Katayama}}]{Fuchizaki2008}%
  \BibitemOpen
  \bibfield  {author} {\bibinfo {author} {\bibfnamefont {K.}~\bibnamefont
  {Fuchizaki}}, \bibinfo {author} {\bibfnamefont {T.}~\bibnamefont
  {Nakamichi}}, \bibinfo {author} {\bibfnamefont {H.}~\bibnamefont {Saitoh}}, \
  and\ \bibinfo {author} {\bibfnamefont {Y.}~\bibnamefont {Katayama}},\
  }\bibfield  {title} {\enquote {\bibinfo {title} {Equation of state of
  hexagonal boron nitride},}\ }\href {\doibase 10.1016/j.ssc.2008.09.031}
  {\bibfield  {journal} {\bibinfo  {journal} {Solid State Communications}\
  }\textbf {\bibinfo {volume} {148}},\ \bibinfo {pages} {390--394} (\bibinfo
  {year} {2008})}\BibitemShut {NoStop}%
\bibitem [{\citenamefont {Constantinescu}\ \emph {et~al.}(2013)\citenamefont
  {Constantinescu}, \citenamefont {Kuc},\ and\ \citenamefont
  {Heine}}]{Constantinescu2013}%
  \BibitemOpen
  \bibfield  {author} {\bibinfo {author} {\bibfnamefont {G.}~\bibnamefont
  {Constantinescu}}, \bibinfo {author} {\bibfnamefont {A.}~\bibnamefont {Kuc}},
  \ and\ \bibinfo {author} {\bibfnamefont {T.}~\bibnamefont {Heine}},\
  }\bibfield  {title} {\enquote {\bibinfo {title} {Stacking in bulk and bilayer
  hexagonal boron nitride},}\ }\href {\doibase
  http://dx.doi.org/10.1103/PhysRevLett.111.036104} {\bibfield  {journal}
  {\bibinfo  {journal} {Phys. Rev. Lett.}\ }\textbf {\bibinfo {volume} {111}},\
  \bibinfo {pages} {036104} (\bibinfo {year} {2013})}\BibitemShut {NoStop}%
\bibitem [{\citenamefont {Popov}\ \emph {et~al.}(2012)\citenamefont {Popov},
  \citenamefont {Lebedeva}, \citenamefont {Knizhnik}, \citenamefont {Lozovik},\
  and\ \citenamefont {Potapkin}}]{Popov2012}%
  \BibitemOpen
  \bibfield  {author} {\bibinfo {author} {\bibfnamefont {A.~M.}\ \bibnamefont
  {Popov}}, \bibinfo {author} {\bibfnamefont {I.~V.}\ \bibnamefont {Lebedeva}},
  \bibinfo {author} {\bibfnamefont {A.~A.}\ \bibnamefont {Knizhnik}}, \bibinfo
  {author} {\bibfnamefont {Yu.~E.}\ \bibnamefont {Lozovik}}, \ and\ \bibinfo
  {author} {\bibfnamefont {B.~V.}\ \bibnamefont {Potapkin}},\ }\bibfield
  {title} {\enquote {\bibinfo {title} {Barriers to motion and rotation of
  graphene layers based on measurements of shear mode frequencies},}\ }\href
  {\doibase 10.1016/j.cplett.2012.03.082} {\bibfield  {journal} {\bibinfo
  {journal} {Chem. Phys. Lett.}\ }\textbf {\bibinfo {volume} {536}},\ \bibinfo
  {pages} {82--86} (\bibinfo {year} {2012})}\BibitemShut {NoStop}%
\bibitem [{\citenamefont {Kolmogorov}\ and\ \citenamefont
  {Crespi}(2005)}]{Kolmogorov2005}%
  \BibitemOpen
  \bibfield  {author} {\bibinfo {author} {\bibfnamefont {A.~N.}\ \bibnamefont
  {Kolmogorov}}\ and\ \bibinfo {author} {\bibfnamefont {V.~H.}\ \bibnamefont
  {Crespi}},\ }\bibfield  {title} {\enquote {\bibinfo {title}
  {Registry-dependent interlayer potential for graphitic systems},}\ }\href
  {\doibase 10.1103/PhysRevB.71.235415} {\bibfield  {journal} {\bibinfo
  {journal} {Phys. Rev. B}\ }\textbf {\bibinfo {volume} {71}},\ \bibinfo
  {pages} {235415} (\bibinfo {year} {2005})}\BibitemShut {NoStop}%
\bibitem [{\citenamefont {Reguzzoni}\ \emph {et~al.}(2012)\citenamefont
  {Reguzzoni}, \citenamefont {Fasolino}, \citenamefont {Molinari},\ and\
  \citenamefont {Righi}}]{Reguzzoni2012}%
  \BibitemOpen
  \bibfield  {author} {\bibinfo {author} {\bibfnamefont {M.}~\bibnamefont
  {Reguzzoni}}, \bibinfo {author} {\bibfnamefont {A.}~\bibnamefont {Fasolino}},
  \bibinfo {author} {\bibfnamefont {E.}~\bibnamefont {Molinari}}, \ and\
  \bibinfo {author} {\bibfnamefont {M.~C.}\ \bibnamefont {Righi}},\ }\bibfield
  {title} {\enquote {\bibinfo {title} {Potential energy surface for graphene on
  graphene: \textit{Ab initio} derivation, analytical description, and
  microscopic interpretation},}\ }\href {\doibase 10.1103/PhysRevB.86.245434}
  {\bibfield  {journal} {\bibinfo  {journal} {Phys. Rev. B}\ }\textbf {\bibinfo
  {volume} {86}},\ \bibinfo {pages} {245434} (\bibinfo {year}
  {2012})}\BibitemShut {NoStop}%
\bibitem [{\citenamefont {Aoki}\ and\ \citenamefont
  {Amawashi}(2007)}]{Aoki2007}%
  \BibitemOpen
  \bibfield  {author} {\bibinfo {author} {\bibfnamefont {M.}~\bibnamefont
  {Aoki}}\ and\ \bibinfo {author} {\bibfnamefont {H.}~\bibnamefont
  {Amawashi}},\ }\bibfield  {title} {\enquote {\bibinfo {title} {Dependence of
  band structures on stacking and field in layered graphene},}\ }\href
  {\doibase 10.1016/j.ssc.2007.02.013} {\bibfield  {journal} {\bibinfo
  {journal} {Solid State Communications}\ }\textbf {\bibinfo {volume} {142}},\
  \bibinfo {pages} {123--127} (\bibinfo {year} {2007})}\BibitemShut {NoStop}%
\bibitem [{\citenamefont {Ershova}\ \emph {et~al.}(2010)\citenamefont
  {Ershova}, \citenamefont {Lillestolen},\ and\ \citenamefont
  {Bichoutskaia}}]{Ershova2010}%
  \BibitemOpen
  \bibfield  {author} {\bibinfo {author} {\bibfnamefont {O.~V.}\ \bibnamefont
  {Ershova}}, \bibinfo {author} {\bibfnamefont {T.~C.}\ \bibnamefont
  {Lillestolen}}, \ and\ \bibinfo {author} {\bibfnamefont {E.}~\bibnamefont
  {Bichoutskaia}},\ }\bibfield  {title} {\enquote {\bibinfo {title} {Study of
  polycyclic aromatic hydrocarbons adsorbed on graphene using density
  functional theory with empirical dispersion correction},}\ }\href {\doibase
  10.1039/C000370K} {\bibfield  {journal} {\bibinfo  {journal} {Phys. Chem.
  Chem. Phys.}\ }\textbf {\bibinfo {volume} {12}},\ \bibinfo {pages}
  {6483--6491} (\bibinfo {year} {2010})}\BibitemShut {NoStop}%
\bibitem [{\citenamefont {Lebedeva}\ \emph {et~al.}(2010)\citenamefont
  {Lebedeva}, \citenamefont {Knizhnik}, \citenamefont {Popov}, \citenamefont
  {Ershova}, \citenamefont {Lozovik},\ and\ \citenamefont
  {Potapkin}}]{Lebedeva2010}%
  \BibitemOpen
  \bibfield  {author} {\bibinfo {author} {\bibfnamefont {I.~V.}\ \bibnamefont
  {Lebedeva}}, \bibinfo {author} {\bibfnamefont {A.~A.}\ \bibnamefont
  {Knizhnik}}, \bibinfo {author} {\bibfnamefont {A.~M.}\ \bibnamefont {Popov}},
  \bibinfo {author} {\bibfnamefont {O.~V.}\ \bibnamefont {Ershova}}, \bibinfo
  {author} {\bibfnamefont {Yu.~E.}\ \bibnamefont {Lozovik}}, \ and\ \bibinfo
  {author} {\bibfnamefont {B.~V.}\ \bibnamefont {Potapkin}},\ }\bibfield
  {title} {\enquote {\bibinfo {title} {Fast diffusion of graphene flake on
  graphene layer},}\ }\href {\doibase 10.1103/PhysRevB.82.155460} {\bibfield
  {journal} {\bibinfo  {journal} {Phys. Rev. B}\ }\textbf {\bibinfo {volume}
  {82}},\ \bibinfo {pages} {155460} (\bibinfo {year} {2010})}\BibitemShut
  {NoStop}%
\bibitem [{\citenamefont {Lebedeva}\ \emph
  {et~al.}(2011{\natexlab{b}})\citenamefont {Lebedeva}, \citenamefont
  {Knizhnik}, \citenamefont {Popov}, \citenamefont {Ershova}, \citenamefont
  {Lozovik},\ and\ \citenamefont {Potapkin}}]{Lebedeva2011a}%
  \BibitemOpen
  \bibfield  {author} {\bibinfo {author} {\bibfnamefont {I.~V.}\ \bibnamefont
  {Lebedeva}}, \bibinfo {author} {\bibfnamefont {A.~A.}\ \bibnamefont
  {Knizhnik}}, \bibinfo {author} {\bibfnamefont {A.~M.}\ \bibnamefont {Popov}},
  \bibinfo {author} {\bibfnamefont {O.~V.}\ \bibnamefont {Ershova}}, \bibinfo
  {author} {\bibfnamefont {Yu.~E.}\ \bibnamefont {Lozovik}}, \ and\ \bibinfo
  {author} {\bibfnamefont {B.~V.}\ \bibnamefont {Potapkin}},\ }\bibfield
  {title} {\enquote {\bibinfo {title} {Diffusion and drift of graphene flake on
  graphite surface},}\ }\href {\doibase 10.1063/1.3557819} {\bibfield
  {journal} {\bibinfo  {journal} {J. Chem. Phys.}\ }\textbf {\bibinfo {volume}
  {134}},\ \bibinfo {pages} {104505} (\bibinfo {year}
  {2011}{\natexlab{b}})}\BibitemShut {NoStop}%
\bibitem [{\citenamefont {Dion}\ \emph {et~al.}(2004)\citenamefont {Dion},
  \citenamefont {Rydberg}, \citenamefont {Schr\"{o}der}, \citenamefont
  {Langreth},\ and\ \citenamefont {Lundqvist}}]{Dion2004}%
  \BibitemOpen
  \bibfield  {author} {\bibinfo {author} {\bibfnamefont {M.}~\bibnamefont
  {Dion}}, \bibinfo {author} {\bibfnamefont {H.}~\bibnamefont {Rydberg}},
  \bibinfo {author} {\bibfnamefont {E.}~\bibnamefont {Schr\"{o}der}}, \bibinfo
  {author} {\bibfnamefont {D.~C.}\ \bibnamefont {Langreth}}, \ and\ \bibinfo
  {author} {\bibfnamefont {B.~I.}\ \bibnamefont {Lundqvist}},\ }\bibfield
  {title} {\enquote {\bibinfo {title} {Van der waals density functional for
  general geometries},}\ }\href {\doibase 10.1103/PhysRevLett.92.246401}
  {\bibfield  {journal} {\bibinfo  {journal} {Phys. Rev. Lett.}\ }\textbf
  {\bibinfo {volume} {92}},\ \bibinfo {pages} {246401} (\bibinfo {year}
  {2004})}\BibitemShut {NoStop}%
\bibitem [{\citenamefont {Marom}\ \emph {et~al.}(2010)\citenamefont {Marom},
  \citenamefont {Bernstein}, \citenamefont {Garel}, \citenamefont {Tkatchenko},
  \citenamefont {Joselevich}, \citenamefont {Kronik},\ and\ \citenamefont
  {Hod}}]{Marom2010}%
  \BibitemOpen
  \bibfield  {author} {\bibinfo {author} {\bibfnamefont {N.}~\bibnamefont
  {Marom}}, \bibinfo {author} {\bibfnamefont {J.}~\bibnamefont {Bernstein}},
  \bibinfo {author} {\bibfnamefont {J.}~\bibnamefont {Garel}}, \bibinfo
  {author} {\bibfnamefont {A.}~\bibnamefont {Tkatchenko}}, \bibinfo {author}
  {\bibfnamefont {E.}~\bibnamefont {Joselevich}}, \bibinfo {author}
  {\bibfnamefont {L.}~\bibnamefont {Kronik}}, \ and\ \bibinfo {author}
  {\bibfnamefont {O.}~\bibnamefont {Hod}},\ }\bibfield  {title} {\enquote
  {\bibinfo {title} {Stacking and registry effects in layered materials: The
  case of hexagonal boron nitride},}\ }\href {\doibase
  10.1103/PhysRevLett.105.046801} {\bibfield  {journal} {\bibinfo  {journal}
  {Phys. Rev. Lett.}\ }\textbf {\bibinfo {volume} {105}},\ \bibinfo {pages}
  {046801} (\bibinfo {year} {2010})}\BibitemShut {NoStop}%
\bibitem [{\citenamefont {Gao}\ and\ \citenamefont
  {Tkatchenko}(2015)}]{Gao2015}%
  \BibitemOpen
  \bibfield  {author} {\bibinfo {author} {\bibfnamefont {W.}~\bibnamefont
  {Gao}}\ and\ \bibinfo {author} {\bibfnamefont {A.}~\bibnamefont
  {Tkatchenko}},\ }\bibfield  {title} {\enquote {\bibinfo {title} {Sliding
  mechanisms in multilayered hexagonal boron nitride and graphene: The effects
  of directionality, thickness, and sliding constraints},}\ }\href {\doibase
  10.1103/PhysRevLett.114.096101} {\bibfield  {journal} {\bibinfo  {journal}
  {Phys. Rev. Lett.}\ }\textbf {\bibinfo {volume} {114}},\ \bibinfo {pages}
  {096101} (\bibinfo {year} {2015})}\BibitemShut {NoStop}%
\bibitem [{\citenamefont {Pulay}(1980)}]{Pulay1980}%
  \BibitemOpen
  \bibfield  {author} {\bibinfo {author} {\bibfnamefont {P.}~\bibnamefont
  {Pulay}},\ }\bibfield  {title} {\enquote {\bibinfo {title} {Convergence
  acceleration of iterative sequences. the case of scf iteration},}\ }\href
  {\doibase 10.1016/0009-2614(80)80396-4} {\bibfield  {journal} {\bibinfo
  {journal} {Chem. Phys. Lett.}\ }\textbf {\bibinfo {volume} {73}},\ \bibinfo
  {pages} {393--398} (\bibinfo {year} {1980})}\BibitemShut {NoStop}%
\bibitem [{\citenamefont {Blakslee}\ \emph {et~al.}(1970)\citenamefont
  {Blakslee}, \citenamefont {Proctor}, \citenamefont {Seldin}, \citenamefont
  {Spence},\ and\ \citenamefont {Weng}}]{Blakslee1970}%
  \BibitemOpen
  \bibfield  {author} {\bibinfo {author} {\bibfnamefont {O.~L.}\ \bibnamefont
  {Blakslee}}, \bibinfo {author} {\bibfnamefont {D.~G.}\ \bibnamefont
  {Proctor}}, \bibinfo {author} {\bibfnamefont {E.~J.}\ \bibnamefont {Seldin}},
  \bibinfo {author} {\bibfnamefont {G.~B.}\ \bibnamefont {Spence}}, \ and\
  \bibinfo {author} {\bibfnamefont {T.}~\bibnamefont {Weng}},\ }\bibfield
  {title} {\enquote {\bibinfo {title} {Elastic constants of
  compression-annealed pyrolytic graphite},}\ }\href {\doibase
  10.1063/1.1659428} {\bibfield  {journal} {\bibinfo  {journal} {J. Appl.
  Phys.}\ }\textbf {\bibinfo {volume} {41}},\ \bibinfo {pages} {3373--3382}
  (\bibinfo {year} {1970})}\BibitemShut {NoStop}%
\bibitem [{\citenamefont {Lee}\ \emph {et~al.}(2008)\citenamefont {Lee},
  \citenamefont {Wei}, \citenamefont {Kysar},\ and\ \citenamefont
  {Hone}}]{Lee2008}%
  \BibitemOpen
  \bibfield  {author} {\bibinfo {author} {\bibfnamefont {C.}~\bibnamefont
  {Lee}}, \bibinfo {author} {\bibfnamefont {X.}~\bibnamefont {Wei}}, \bibinfo
  {author} {\bibfnamefont {J.~W.}\ \bibnamefont {Kysar}}, \ and\ \bibinfo
  {author} {\bibfnamefont {J.}~\bibnamefont {Hone}},\ }\bibfield  {title}
  {\enquote {\bibinfo {title} {Measurement of the elastic properties and
  intrinsic strength of monolayer graphene},}\ }\href {\doibase
  10.1126/science.1157996} {\bibfield  {journal} {\bibinfo  {journal}
  {Science}\ }\textbf {\bibinfo {volume} {321}},\ \bibinfo {pages} {385--388}
  (\bibinfo {year} {2008})}\BibitemShut {NoStop}%
\bibitem [{\citenamefont {Yang}\ \emph {et~al.}(2011)\citenamefont {Yang},
  \citenamefont {Barsoum},\ and\ \citenamefont {Rethinam}}]{Yang2011}%
  \BibitemOpen
  \bibfield  {author} {\bibinfo {author} {\bibfnamefont {B.}~\bibnamefont
  {Yang}}, \bibinfo {author} {\bibfnamefont {M.~W.}\ \bibnamefont {Barsoum}}, \
  and\ \bibinfo {author} {\bibfnamefont {R.~M.}\ \bibnamefont {Rethinam}},\
  }\bibfield  {title} {\enquote {\bibinfo {title} {Nanoscale continuum
  calculation of basal dislocation core structures in graphite},}\ }\href
  {\doibase 10.1080/14786435.2010.539189} {\bibfield  {journal} {\bibinfo
  {journal} {Philosophical Magazine}\ }\textbf {\bibinfo {volume} {91}},\
  \bibinfo {pages} {1441--1463} (\bibinfo {year} {2011})}\BibitemShut {NoStop}%
\bibitem [{\citenamefont {Ni}\ \emph {et~al.}(2008)\citenamefont {Ni},
  \citenamefont {Yu}, \citenamefont {Lu}, \citenamefont {Wang}, \citenamefont
  {Feng},\ and\ \citenamefont {Shen}}]{Ni2008}%
  \BibitemOpen
  \bibfield  {author} {\bibinfo {author} {\bibfnamefont {Z.~H.}\ \bibnamefont
  {Ni}}, \bibinfo {author} {\bibfnamefont {T.}~\bibnamefont {Yu}}, \bibinfo
  {author} {\bibfnamefont {Y.~H.}\ \bibnamefont {Lu}}, \bibinfo {author}
  {\bibfnamefont {Y.~Y.}\ \bibnamefont {Wang}}, \bibinfo {author}
  {\bibfnamefont {Y.~P.}\ \bibnamefont {Feng}}, \ and\ \bibinfo {author}
  {\bibfnamefont {Z.~X.}\ \bibnamefont {Shen}},\ }\bibfield  {title} {\enquote
  {\bibinfo {title} {Uniaxial strain on graphene: Raman spectroscopy study and
  band-gap opening},}\ }\href {\doibase 10.1021/nn800459e} {\bibfield
  {journal} {\bibinfo  {journal} {ACS Nano}\ }\textbf {\bibinfo {volume} {2}},\
  \bibinfo {pages} {2301--2305} (\bibinfo {year} {2008})}\BibitemShut {NoStop}%
\bibitem [{\citenamefont {Li}\ \emph {et~al.}(2008)\citenamefont {Li},
  \citenamefont {Wang}, \citenamefont {Zhang}, \citenamefont {Lee},\ and\
  \citenamefont {Dai}}]{Li2008}%
  \BibitemOpen
  \bibfield  {author} {\bibinfo {author} {\bibfnamefont {X.}~\bibnamefont
  {Li}}, \bibinfo {author} {\bibfnamefont {X.}~\bibnamefont {Wang}}, \bibinfo
  {author} {\bibfnamefont {L.}~\bibnamefont {Zhang}}, \bibinfo {author}
  {\bibfnamefont {S.}~\bibnamefont {Lee}}, \ and\ \bibinfo {author}
  {\bibfnamefont {H.}~\bibnamefont {Dai}},\ }\bibfield  {title} {\enquote
  {\bibinfo {title} {Chemically derived, ultrasmooth graphene nanoribbon
  semiconductors},}\ }\href {\doibase 10.1126/science.1150878} {\bibfield
  {journal} {\bibinfo  {journal} {Science}\ }\textbf {\bibinfo {volume}
  {319}},\ \bibinfo {pages} {1229--1232} (\bibinfo {year} {2008})}\BibitemShut
  {NoStop}%
\end{thebibliography}%

\end{document}